\documentclass[iop,apj,tighten,revtex4]{emulateapj}
\usepackage{color}
\usepackage{natbib}
\usepackage{amssymb}
\usepackage{amsmath}

\begin{document}

\title[Nova Light Curves from SMEI - II. The extended catalog]{Nova Light Curves From The Solar Mass Ejection Imager (SMEI) - II. The extended catalog}
\author{R. Hounsell$^{1,2}$, M. J. Darnley$^3$, M. F. Bode$^3$, D. J. Harman$^3$, F. Surina$^{3,4}$,\\ S. Starrfield$^5$,  D. L. Holdsworth$^{6}$, D. Bewsher$^{7}$, P. P. Hick$^8$, B. V. Jackson$^9$, \\A. Buffington$^9$, J. M. Clover$^9$, A. W. Shafter$^{10}$}

\affiliation{$^1$Space Telescope Science Institute, 3700 San Martin Drive, Baltimore, MD, 21218, USA}
\affiliation{$^2$Department of Astronomy, University of Illinois Urbana Champaign, 1002 W Green St, Urbana, IL, 61801, USA}
\affiliation{$^3$Astrophysics Research Institute, Liverpool John Moores University, IC2 Liverpool Science Park, Liverpool, L3 5RF, UK}
\affiliation{$^4$Department of Science and Technology, Chiangrai Rajabhat University, 80 Moo 9, Bandu, Muang, Chiangrai, 57100, Thailand}
\affiliation{$^5$School of Earth and Space Exploration, Arizona State University, PO Box 871404, Tempe, AZ 85287-1404, USA}
\affiliation{$^6$Astrophysics Group, Keele University, Staffordshire ST5 5BG, UK}
\affiliation{$^7$Jeremiah Horrocks Institute, University of Central Lancashire, Preston PR1 2HE, UK}
\affiliation{$^8$CAIDA UCSD/SDSC 9500 Gilman Dr. Mail Stop 0505 La Jolla, CA 92093-0505, USA}
\affiliation{$^9$Center for Astrophysics and Space Science, University of California, San Diego, 9500 Gilman Drive 0424, La Jolla, CA 92093-0424, USA}
\affiliation{$^{10}$Department of Astronomy, San Diego State University, San Diego, CA 92182, USA}

\begin{abstract}
We present the results from observing nine Galactic novae in eruption with the Solar Mass Ejection Imager (SMEI) between 2004 and 2009. While many of these novae reached peak magnitudes that were either at or approaching the detection limits of SMEI, we were still able to produce light curves that in many cases contained more data at and around the initial rise, peak, and decline than those found in other variable star catalogs. For each nova, we obtained a peak time, maximum magnitude, and for several an estimate of the decline time ($t_{\rm 2}$). Interestingly, although of lower quality than those found in \citet{Hounsell}, two of the light curves may indicate the presence of a pre-maximum halt. In addition the high cadence of the SMEI instrument has allowed the detection of low amplitude variations in at least one of the nova light curves.\end{abstract}


\section{Introduction}
\label{intro}
Classical Novae (CNe) belong to the Cataclysmic Variable (CV) class of objects. These are interacting close binary systems in which mass is transferred from a donor star to the surface of an accreting compact companion leading to a variety of behaviors, the most noticeable of which is an eruption \citep[see][for reviews]{2008clno.book.....B, 2010AN....331..160B}. 

A typical CN system consists of a white dwarf (WD) primary and a cooler lower mass main sequence secondary star \citep[spectral type ranging from F to M;][]{2012ApJ...746...61D}. As the secondary evolves it fills its Roche lobe allowing hydrogen rich material to transfer at a rate of 10$^{-11}-10^{-9}$ M$_{\odot}$ yr$^{-1}$ via the inner Lagrangian point (L1) toward the WD and form an accretion disc. Material is deposited from the disk onto the surface of the WD, causing the bottom of the accreted layer on the WD to become electron degenerate. Within this degenerate layer hydrogen burning ensues leading to a thermonuclear runaway (TNR) and ultimately the CN eruption. The total radiant output of a single CN eruption is in the range of 10$^{45} - 10^{46}$ erg with TN-outburst amplitudes of approximately 10$-$20 magnitudes, and absolute {\it V}-band magnitudes of $M_\mathrm{V}= -10.7$ at maximum for the fastest and most luminous CNe \citep[][and references therein]{Shafter09}. The resulting energy released is sufficient to expel the accreted envelope and drive mass loss \citep[$10^{-5} - 10^{-4}~M_{\odot}$,][] {2011MNRAS.411..162G} at velocities of a few hundred to several thousand km s$^{-1}$. All CNe are thought to have repeat eruptions \citep[although for some this may take up to 10$^{4}/10^{5}$ years,][]{2005ASPC..330..265H} however, a CN observed in TN-outburst more than once is reclassified as a recurrent nova (RN); observed recurrence times range from 1-100 years \citep{2014A&A...563L...9D,2015A&A...580A..45D}, or even as short as six months \citep{2015A&A...582L...8H}, with TN-outburst amplitudes typically smaller than CNe ($\approx$ 5-6 magnitudes if the secondary is giant, 10-15 if a main sequence star). The basic triggering and explosion mechanism of a RN is the same as for a CN, but there are some distinct differences in the physical properties of this subgroup. In order to reconcile the short quiescence period of a RN with a TNR the WDs within these systems are believed to be hotter, more massive \citep[close to the][limit]{1931ApJ....74...81C}, and have higher mass accretion rates ($\sim$10$^{-8}-10^{-7}M_{\odot}$ yr$^{-1}$) than CNe \citep[see][]{2005ApJ...623..398Y, 2013ApJ...777..136W}. The majority of RN systems also harbor evolved secondary stars rather than the typical main sequence star of a CN system \citep[see, for example, ][]{2012ApJ...746...61D}.

As with most transients, CNe are traditionally classified according to their photometric and spectroscopic properties. Each nova has its own unique optical light curve. However, they do share a common idealized nova light curve \citep{1960stat.conf..585M}. Most novae tend to rise rapidly to peak within one to three days. Due to the transient nature of a nova this initial rise has rarely been observed well enough to establish any classification regime, but the subsequent decline from maximum has. Novae tend to be classified according to the number of days that they take to decline {\it n} magnitudes from maximum, thus they are divided into ``speed classes''. These classes were first introduced by \citet{1957gano.book.....G}, and often declines of two or three magnitudes are quoted ($t_{2}$ and $t_{3}$ respectively).

During the initial rise, novae are thought to experience a pre-maximum halt (PMH) approximately one to two magnitudes below maximum \citep{1960stat.conf..585M}. The duration of this halt appears to be related to the speed class of the nova, lasting a few hours for fast novae and a few days for slow novae. Until recently the PMH had only been observed for a few slow novae such as DQ Her, V450 Cyg \citep{1964gano.book.....P}, V723 Cas, V463 Sct \citep[][who present a new interpretation of the observed long PMHs]{2004ApJ...612L..57H}, and V5558 Sgr \citep{2011PASJ...63..911T}  with no strong evidence to suggest that it was also present for the faster speed classes. \citet[][hereafter Paper~I]{Hounsell} however, found from SMEI observations that for all very fast/fast novae (classical and recurrent) studied in detail in that paper, a PMH was detected with a duration consistent with their speed class. The PMHs of the very fast novae observed in Paper~I consisted of 3-4 SMEI data points in which a temporary reversal of the light curve was suggested and its duration was calculated as the time between the first and third change in gradient of the slope. The PMH of the one fast nova detected (V1280 Sco) was represented by a plateau in the  light curve and made up of 7 SMEI data points. Here the PMH duration was taken as the time between the first and second change of gradient. If a slower nova were to have been detected within the SMEI data we might expect a similar plateau.

Although there is now evidence for the existence of this phase of evolution in the nova light curve, no physical explanation for the PMH is currently accepted although several theories exist. \citet{2012AJ....144...98W} suggests that a PMH lasting a day or more may be caused by an enhancement of mass loss from the secondary star, which then dominates over the initial WD ejecta. This changeover in the dominance of the two mass loss sources could be the cause of the halt. Other work by \citet{Hillman} has used nova evolution simulations (via a hydrodynamic Lagrangian code) to create detailed light curves, and find that halts of long or short duration occur naturally within the burst. Examining the evolution of nova effective temperature ($T_{\rm eff}$), luminosity, and radius they find that at a point just before expansion and mass loss occur, that the $T_{\rm eff}$ deceases and the rising luminosity halts. Their work also found that this halt would in many cases be accompanied by a dip in the total luminosity. Such a dip is due to a temporary drop in the energy flux as convection in the expanding and thinning envelope ceases to be efficient near the envelope surface. Because of the decreasing opacity of the envelope, radiation soon dominates over convection thus reversing the dip. Work by \citet{Hillman} and \citet{2012AJ....144...98W} are both WD mass dependent. 

The majority of Galactic novae are still discovered by the amateur community. Over the past century various surveys have attempted to measure the Galactic rate, however results are greatly affected by temporal and spatial coverage, selection effects, and interstellar extinction.  Based on an extrapolation of the observed nova rates in the solar neighborhood (assumed complete to m$<$2), \citet{1997ApJ...487..226S} estimated a Galactic nova rate of approximately 35~yr$^{-1}$, this is supported by \citet{2006MNRAS.369..257D} with a value of 34$_{-12}^{+15}$ yr$^{-1}$. Of these an average of roughly one CN per year has been {\it observed} to reach $m_{\rm V}$ = 8 or brighter \citep[see Figure 2 of][]{2002AIPC..637..462S}. It is well known that novae with the highest peak bolometric luminosity fade the most rapidly and are thus often missed; historical observations are therefore clearly incomplete at $m_{V}$ = 8 and the actual number of novae reaching this brightness is expected to be significantly higher \citep[$\sim$5 yr $^{-1}$, see][and discussion in Paper~I]{2002AIPC..637..462S, WarnerB}.\\

With the advent of all-sky imaging facilities, both ground and space-based, there is new hope for detecting a more complete sample of Galactic novae. Although the detection of transient events may not be the original science objective of these missions, their archives hold a wealth of data on many events. These observations may contain great detail and provide data on many previously poorly examined and understood phases of evolution. Examination of these archives is therefore exceptionally important. One such space-based all-sky mission is the Solar Mass Ejection Imager (SMEI), its usefulness in observing novae was well documented in Paper I, \citet{2010ATel.2558....1H, 2011ATel.3373....1H, 2012IAUS..285...91H}, \citet{2014AJ....147..107S}, and \citet{2013arXiv1303.2711D}.

SMEI is a high precision white light differential photometer \citep{2006ApJ...637..880B, 2007SPIE.6689E...8B} based on board the Coriolis solar satellite. SMEI was in operation from January 2003 \citep{2003SoPh..217..319E, 2004SoPh..225..177J} until September 2011. The instrument consists of three baffled CCD cameras each with a 60$^{\circ} \times$ 3$^{\circ}$ field of view, combining to sweep out nearly the entire sky with each 102 minute orbit of the spacecraft \citep{2007SPIE.6689E...9H}. The peak throughput of the instrument is at approximately 700~nm with a FWHM $\sim$~300 nm. SMEI is able to reliably detect brightness changes in point sources down to at least 8$^{th}$ magnitude.

SMEI was originally designed to map out large-scale variations in heliospheric electron densities by observing the Thomson-scattered sunlight from solar wind electrons \citep{2004SoPh..225..177J}. In order to isolate the faint Thomson-scattered sunlight, the much larger white-light contributions from the zodiacal dust cloud, the sidereal background, and individual point sources (bright stars and planets) were determined and removed \citep[see][for further details]{2007SPIE.6689E...9H}. Thus, brightness determination of point sources is a routine step in the SMEI data analysis, and as such has led to the production of detailed light curves for many bright objects including variables \citep[see][and Paper I for example]{2006ApJ...637..880B, 2007ASPC..366...39S, 2007MNRAS.382L..48T, 2008CoAst.157...92T, 2008A&A...492..167T, 2008A&A...483L..43T, 2011AAS...21725702C, 2011MNRAS.411..162G, 2014AJ....147..107S}. 

This paper presents results from the examination of nine additional Galactic novae observed by SMEI. These novae are fainter than the four presented in Paper I and indicate the limit of SMEI's ability to detect brightness changes at fainter magnitudes and in crowded regions. In Section~\ref{data_analysis} we explain how data were obtained and analyzed. Section~\ref{results} presents our results, and a discussion along with our conclusions are provided in Section~\ref{disscussion_conculsion}.
                                                                                                                                                                                                                                                                                                                                                                                                                                                                                                                                                                                                                                                                                     
\section{Data Analysis}
\label{data_analysis}
The SMEI database contains a catalog with the names, co-ordinates, and discovery magnitudes of 62 Galactic CNe and 3 RNe, with eruptions dating between 2003 and 2011. The photometry of each nova in this paper was obtained using this catalog and an extended iterative least-squares fit of the point spread function (PSF) as described in \citet{2005SPIE.5901..340H, 2007SPIE.6689E...9H}; and Section 2 of Paper I. Zodiacal and sidereal background light contributions were also considered during the fitting stage. The remaining 51 novae not examined as part of this paper, \citet{2014AJ....147..107S}, or Paper~I were either too faint for detection or resided in fields far too crowded for reliable results to be obtained.

As noted in Paper I, the SMEI PSF has a full width of approximately 1$^{\circ}$ and is highly asymmetric with a ``fish-like" appearance. Due to this large size an object of interest is considered crowded when it lies less than one PSF width from another bright object (typically 6$^{th}$ magnitude or brighter); such crowding within the SMEI data is commonplace. To combat this issue simultaneous fitting of multiple objects can be initiated where possible (stellar separation $>$0.75$^{\circ}$) and contamination of the object of interest reduced.

In order to achieve the best fit and produce the most reliable nova light curve, the surrounding region of each object required assessment for levels of potential contamination due to crowding, cosmic-ray hits, and finally their proximity to the Sunwards and anti-Sunwards masks \citep[see][for a description of these masks]{2006ApJ...637..880B, 2007SPIE.6689E...8B}. If the nova was in a crowded field, a simultaneous fit was conducted and the area used to sample the surrounding stellar region (wing radius) from the PSF centriod reduced. The reduction of the wing radius also reduces errors if the object is located near a masked region. For these reasons multiple light curves were generated for each nova using an {\it auto\_wing}\footnotemark[1] radius and wing radii of 1.2$^{\circ}$, 1.3$^{\circ}$, and 1.4$^{\circ}$. The resulting photometry files were then assessed on several criteria: 1. correlation of the fitted PSF to the model $-$ sample Pearson correlation coefficient (~{\it r}~); 2. the number of pixels used to define the PSF fitting area $-$ often an indictor of masks or image defects (~{\it npsf}~); 3. variation of the background fitted value; 4. deviation of the RA and dec from the catalog position $-$ jumps can be an indication of the fitting of a residual from the poor subtraction of a neighboring star, this becomes more of an issue as the nova fades.

\footnotetext[1]{This is defined by the following equation: {\it auto\_wing} =  $0.1 \times (3 - m_{\rm{SMEI}}$) + 1.4 (degrees), where $m_{\rm{SMEI}}$ is the magnitude listed for the nova within the catalogue and is often the discovery magnitude only. For novae of interest {\it auto\_wing} is approximately $1^{\circ}$.}

Within Paper~I, points possessing {\it r} $\geq$ 0.5 were deemed reliable. However, as the novae presented here are much fainter, a less stringent cut off of 0.4 was found to be acceptable. A 1$\sigma$ threshold was applied to the other selection criteria listed above (in some cases a 3$\sigma$ cut off was applied if the object was detected only a few times by the instrument, or if the initial RA and dec were uncertain).  Using the filtered file which possessed the highest number of ``valid'' points, the flux of the nova was then converted into an unfiltered SMEI apparent magnitude ($m_{\mathrm{SMEI}}$), and its error contribution calculated from photon counting statistics. 

\section{Results}
\label{results}
\begin{figure}
\centering
\fbox{\includegraphics[keepaspectratio=true,  scale=0.65]{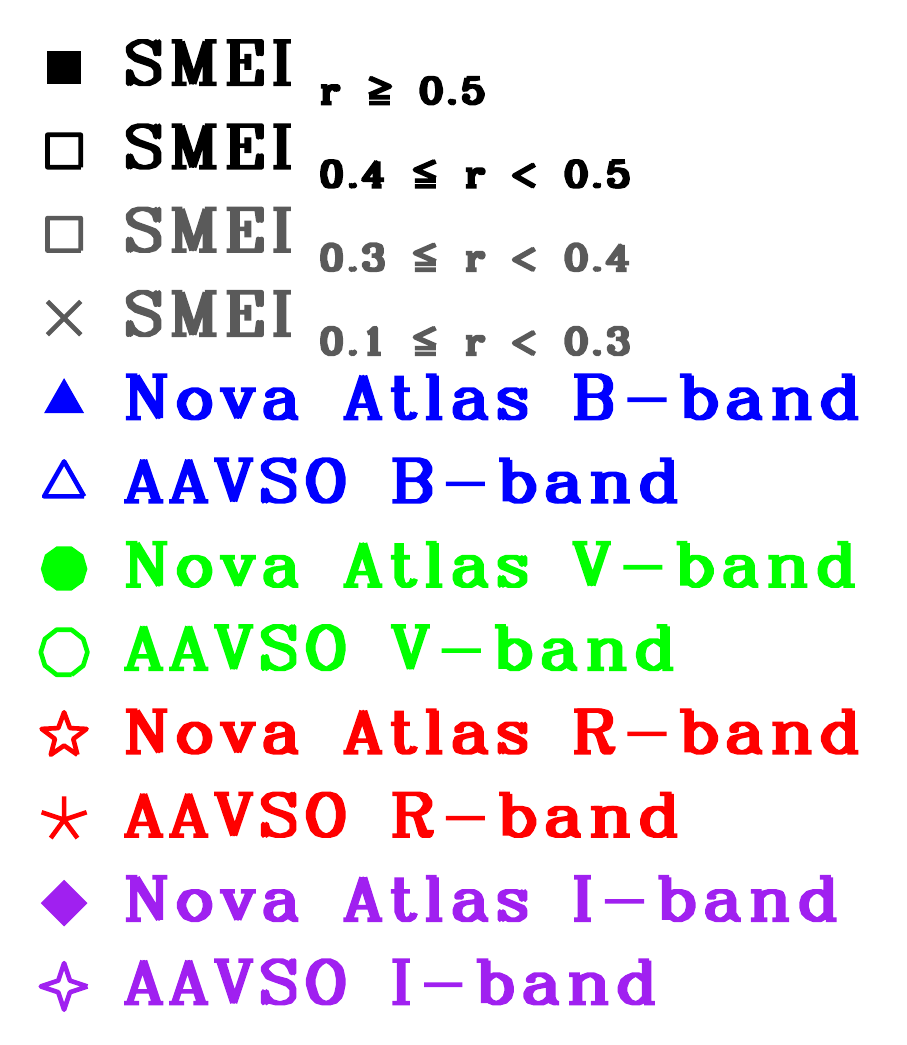}}
\caption{Legend for all data in following figures.}
\label{key}
\end{figure}

As noted above, the novae presented in this paper are much fainter than those given in Paper I, all of which possessed a peak $m_{\rm SMEI}$ $<$ 5.5. The objects within this paper have peak magnitudes between 6 and 8, fainter than this detection is unreliable. The light curves derived, although more noisy than those in Paper~I, have nonetheless provided precise measurements of peak time, eruption magnitude, and in many cases a value for $t_{2}$. Table~\ref{table1} summarizes our main findings, and the individual novae are discussed below. Novae which possess the most interesting data are presented first, with the remaining given in order of eruption date. Where possible a light-curve for each nova has been presented, with a legend for all data given in Figure~\ref{key}. An appendix tabulating the available photometry for each nova examined is included at the end of the paper.

\begin{deluxetable*}{lccccccccc}
\tablewidth{0pt}
\tablecaption{Derived light curve parameters of 14 novae using data from the Solar Mass Ejection Imager. The last five novae listed represent objects reported within Hounsell et al. (2012); Surina et al. (2014) and Paper~I. Novae in this paper tend to be fainter than those in Paper~I, and as such we were unable to obtain in most cases all parameters listed in the table.\label{table1}}
\tablehead{
\colhead{\scriptsize{Name}} & \colhead{\scriptsize{Time of}}  & \colhead{Peak} & \colhead{\scriptsize{$t_{2}$}} & \colhead{\scriptsize{Speed}} & \colhead{\scriptsize{Pre-max}} & \colhead{\scriptsize{Pre-max}} & \colhead{\scriptsize{$\Delta m_{\rm SMEI}$}} & \colhead{\scriptsize{$\Delta t$ from}} & \colhead{\scriptsize{Wing}}\\
& \colhead{\scriptsize{maximum}} & \colhead{\scriptsize{SMEI}} & \colhead{\scriptsize{(days)}}& \colhead{\scriptsize{class\tablenotemark{a}}} & \colhead{\scriptsize{duration}} & \colhead{\scriptsize{mean}} & \colhead{\scriptsize{from halt}}  & \colhead{\scriptsize{halt to}} &  \colhead{\scriptsize{radius}}\\
&\colhead{\scriptsize{(yyyy/mm/dd)}} & \colhead{\scriptsize{magnitude}} &  && \colhead{\scriptsize{(days)\tablenotemark{b}}} &  \colhead{\scriptsize{magnitude}} & \colhead{\scriptsize{to peak\tablenotemark{c}}} &  \colhead{\scriptsize{peak (days)}} & \colhead{\scriptsize{(deg)}}
}
\startdata
\scriptsize{V1187~Sco} & \scriptsize{2004/08/3.77$_{-0.04}^{+0.07}$} & \scriptsize{6.95$\pm$0.04} & \scriptsize{10.10$_{-0.39}^{+0.43}$}\tablenotemark{d}   & \scriptsize{Very fast} & \scriptsize{\nodata} & \scriptsize{\nodata} & \scriptsize{\nodata} & \scriptsize{\nodata} & \scriptsize{1.3} \\
\scriptsize{V2467~Cyg} & \scriptsize{2007/03/16.56$\pm$0.04} & \scriptsize{6.30$\pm$0.03} & \scriptsize{ 5.68$_{-0.53}^{+0.61}$}\tablenotemark{d} &  \scriptsize{Very fast} & \scriptsize{\nodata} & \scriptsize{\nodata} & \scriptsize{\nodata} & \scriptsize{\nodata} & \scriptsize{1.3}\\
\scriptsize{V458~Vul} & \scriptsize{2007/08/13.66$\pm$0.04} & \scriptsize{7.94$\pm$0.07} & \scriptsize{\nodata} & \scriptsize{Fast} & \scriptsize{\nodata} & \scriptsize{\nodata} & \scriptsize{\nodata} & \scriptsize{\nodata} & \scriptsize{{\it auto-wing}}\\
\scriptsize{V597~Pup} & \scriptsize{2007/11/14.68$_{-0.35}^{+0.04}$} & \scriptsize{6.91$\pm$0.04} & \scriptsize{2.81$_{-0.90}^{+1.0}$}\tablenotemark{d} & \scriptsize{Very fast} & \scriptsize{0.21} & \scriptsize{8.07$\pm$ 0.14} & \scriptsize{1.16} & \scriptsize{0.96} & \scriptsize{1.3}\\
\scriptsize{V459~Vul} & \scriptsize{2007/12/28.11$_{-0.07}^{+0.11}$} & \scriptsize{6.59$\pm$0.04} & \scriptsize{$\sim$19.4\tablenotemark{e}} & \scriptsize{Fast} & \scriptsize{\nodata} & \scriptsize{\nodata} & \scriptsize{\nodata} & \scriptsize{\nodata} & \scriptsize{1.2}\\
\scriptsize{V2491~Cyg} & \scriptsize{2008/04/10.89$\pm$0.04} & \scriptsize{7.36$\pm$0.05} & \scriptsize{\nodata} & \scriptsize{Very fast} & \scriptsize{\nodata} & \scriptsize{\nodata} & \scriptsize{\nodata} & \scriptsize{\nodata} & \scriptsize{1.2}\\
\scriptsize{QY~Mus} & \scriptsize{2008/09/28.63$\pm$0.2} & \scriptsize{6.93$\pm$0.04} & \scriptsize{\nodata} & \scriptsize{Moderately fast} & \scriptsize{\nodata} & \scriptsize{\nodata} & \scriptsize{\nodata} & \scriptsize{\nodata} & \scriptsize{1.2}\\
\scriptsize{V5580~Sgr} & \scriptsize{2008/11/30.85} & \scriptsize{7.01$\pm$0.04} & \scriptsize{\nodata} & \scriptsize{\nodata} & \scriptsize{\nodata} & \scriptsize{\nodata} & \scriptsize{\nodata} & \scriptsize{\nodata} & \scriptsize{{\it auto-wing}}\\
\scriptsize{V5583~Sgr} & \scriptsize{2009/08/7.08$\pm$0.04} & \scriptsize{6.94$\pm$0.05} & \scriptsize{\nodata} & \scriptsize{Very fast} & \scriptsize{0.21} & \scriptsize{7.64$\pm$0.03} & \scriptsize{0.71} & \scriptsize{1.10} & \scriptsize{1.4}\\
\tableline
\scriptsize{T~Pyx\tablenotemark{f}} & \scriptsize{2011/05/12.22$\pm$0.04} & \scriptsize{6.33$\pm$0.03} & \scriptsize{\nodata} & \scriptsize{Slow} & \scriptsize{$\sim$10} & \scriptsize{$\sim$8.1} & \scriptsize{$\sim$1.8} & \scriptsize{14.6} & \scriptsize{\nodata} \\
\tableline 
\scriptsize{RS~Oph} & \scriptsize{2006/02/12.94$\pm$0.04} & \scriptsize{3.87$\pm$0.01} & \scriptsize{7.9} & \scriptsize{Very fast} &  \scriptsize{0.14} & \scriptsize{4.50$\pm$0.05} & \scriptsize{0.63} & \scriptsize{0.49} & \scriptsize{1.25}\\
\scriptsize{V1280~Sco} & \scriptsize{2007/02/16.15$\pm$0.04} & \scriptsize{4.00$\pm$0.01} & \scriptsize{21.3} & \scriptsize{Fast} & \scriptsize{0.42} & \scriptsize{5.231$\pm$0.003} & \scriptsize{1.23} & \scriptsize{0.49} & \scriptsize{1.25}\\
\scriptsize{V598~Pup} & \scriptsize{2007/06/6.29$\pm$0.04} & \scriptsize{3.46$\pm$0.01} & \scriptsize{4.3} & \scriptsize{Very fast} & \scriptsize{0.28} & \scriptsize{5.2$\pm$0.1} & \scriptsize{1.74} & \scriptsize{2.19} & \scriptsize{1.4}\\
\scriptsize{KT~Eri} & \scriptsize{2009/11/14.67$\pm$0.04} & \scriptsize{5.42$\pm$0.02} & \scriptsize{6.6} & \scriptsize{Very fast} & \scriptsize{0.14} & \scriptsize{6.04$\pm$0.07} & \scriptsize{0.63} & \scriptsize{0.71} & \scriptsize{1.25}
\enddata

\tablenotetext{a}{V1187 Sco - \citet{Lynch1187}; V2467 Cyg - \citet{2009AAS...21349125L}; V458 Vul -  \citet{2008ApJ...688L..21W}; V597 Pup - \citet{Naik}; V459 Vul -  \citet{Poggiani}; V2491 Cyg - \citet{2011NewA...16..209M} \& Darnley {\it et al.} (2011); QY Mus - \citet{2011ApJS..197...31S}; V5583 Sgr - \citet{2011ApJS..197...31S}}
\tablenotetext{b}{Here the duration of the halt is taken to be the time between the first and third change in gradient of the rising light curve for very fast novae. For fast and moderately fast it is the first and second change.}
\tablenotetext{c}{$\Delta m_{\rm SMEI}$ from halt to peak is calculated using the mean magnitude of the PMH.}
\tablenotetext{d}{Based on a linear fit of the initial decline data from the SMEI nova light curve (see text for further details).}
\tablenotetext{e}{Using linear extrapolation between the first and last data point of the SMEI light curve (see text for further details).}
\tablenotetext{f}{Here all data have been taken from Surina et al. (2014). Within this work the PMH was suggested to have lasted for $\approx$ 10 days with magnitudes varying between 7.7 and 8.5, hence a mean magnitude of 8.1 was quoted in the above table. The time between halt and peak was also taken, in this case, from the end of the PMH phase.}
\end{deluxetable*}

\subsection{V597 Puppis}
\label{subsub:v597Pup}
\begin{figure*}
\centering
\includegraphics[keepaspectratio=true,  scale=0.6]{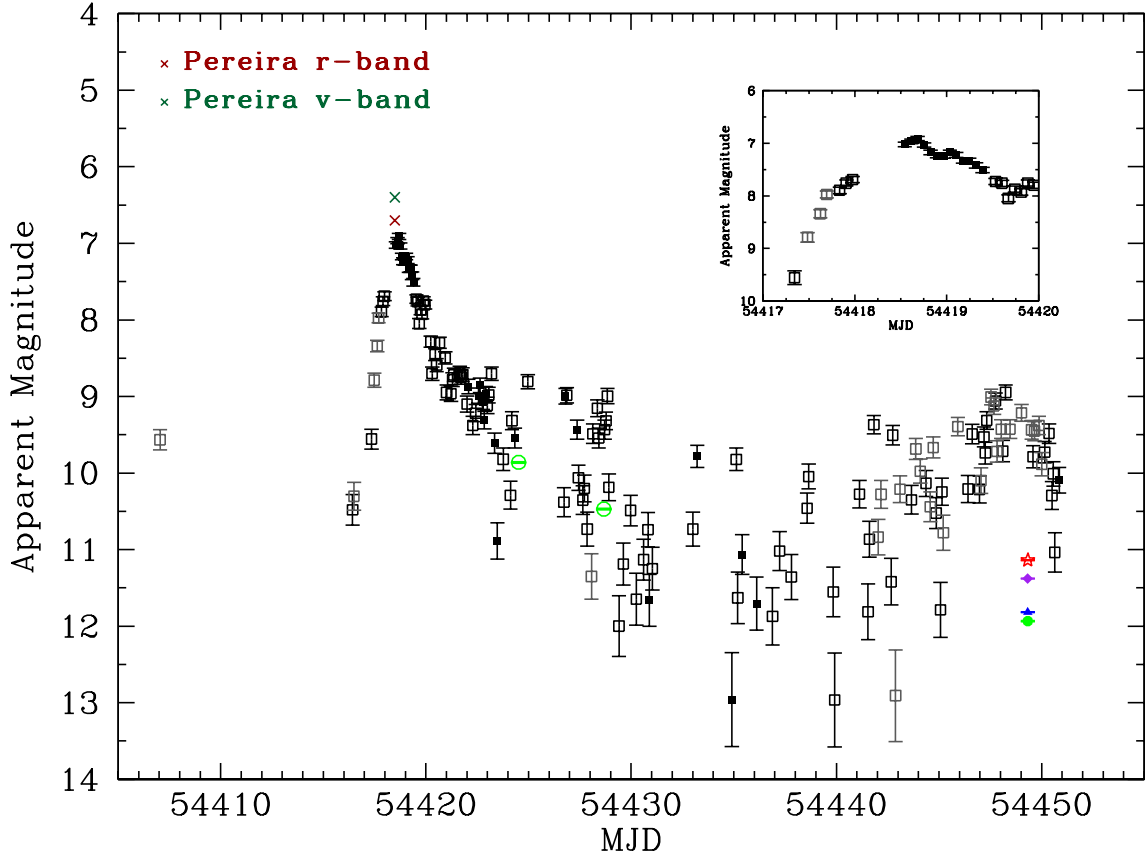}
\caption{Optical light curves of V597~Pup. SMEI data with a lower {\it r} than optimal have been included for completeness. Nova Atlas data which coincide with the SMEI light curve have been included along with data from the AAVSO (see Figure~\ref{key} for data legend). Green and red crosses represent the peak {\it V} and {\it R}-band data given within \citet{Pereira}. The inset shows the final rise, peak, and initial decline of the nova in the SMEI data only, which may show evidence for a PMH around MJD 54417.63.}
\label{v597pup}
\end{figure*}

Nova V597 Pup ($\alpha$ = $08^{h}16^{m}18^{s}\!\!_{.}01,\;$ $\delta$ = $-34^{\circ}15^{\prime}24^{\prime\prime}\!\!_{.}$1; J2000) was discovered in TN-outburst by \citet{Pereira} at a visual magnitude of 7.0 on 2007 November 14.23 UT (MJD 54418.23), reaching a peak visual magnitude of $m_{\rm V}$ = 6.4 on 2007 November 14.48 UT (MJD 54418.48, green cross within Figure~\ref{v597pup}). The nova then declined rapidly with a $t_{\rm 2}$ = 2.5 days \citep{Naik}, making it one of the fastest novae recorded. \citet{Naik} classified the nova as a He/N-type with a WD close to  the Chandrasekhar limit. A pre-eruption detection is found within the Digitized Sky Survey \citep{Pereira} with a source at {\it V} $\sim$ 20, that coincides with the nova position. Continued monitoring of the object by \citet{Warner579} revealed the nova as an intermediate polar (IP) in the orbital period gap ($P_{\rm orb}$ = 2.67 hrs), with a rotational period of 8.7 minutes. Observation of the object a year after eruption also revealed the presence of a deep secondary eclipse caused by the passage of the optically thick accretion disc in front of the irradiated side of the secondary star. The object is considered unique as it is the first CV found to have this deep an eclipse. 

The light curve created using the SMEI data set covers the latter part of the initial rise, peak, and early decline (see Figure~\ref{v597pup}). There does appear to be one data point on 2007 November 7.04 UT (MJD 54407.04) at $m_{\rm SMEI}$= 9.57 $\pm$ 0.13, however the {\it r} of this point is only 0.34 and thus not reliable. All data after this point and up to the initial rise possess an {\it r} less than 0.3 and are therefore omitted from Figure~\ref{v597pup}, and further discussion. Examination of the initial rise hints at the possibility of a PMH starting on 2007 November 13.63 UT (MJD 54417.63) and lasting several hours with a mean $m_{\rm SMEI}$ = 8.07 $\pm$ 0.14 (defined as the mean magnitude over the duration of the halt; quoted error is the rms scatter). As in Paper~I, the duration of the halt for this object is taken to be the time between the first and third changes in gradient of the rising light curve, and is  appropriate for the speed class of the nova. It should be noted however that the first two points of this halt have {\it r} $<$0.4 (actual values are 0.39 and 0.35 respectively, this lower {\it r} is due in part to the limiting magnitude of SMEI) and as such the reality of this PMH can not be confirmed. There is then a gap in the data lasting several hours due to the SMEI weekly calibration, after which the nova is seen to reach its maximum intensity of $m_{\rm SMEI} = 6.91 \pm 0.04$ on 2007 November 14.68$_{-0.35}^{+0.04}$ UT (MJD 54418.68). This coincides better with the {\it R}-band magnitude of 6.7 on 2007 November 14.48 UT (MJD 54418.48) given in \citet[][red cross within Figure~\ref{v597pup}]{Pereira} than the {\it V}-band data quoted above. Note however, it is possible that the peak nova brightness actually occurred in the data gap caused by the weekly calibration. This uncertainty has been included in the quoted time of maximum. The decline of the nova is shown by SMEI to be exceptionally fast, but  given the scatter off the data around 8/9th magnitude a direct measurement of $t_{\rm 2}$ could not be made. However, assuming that the data can be modeled by a simple linear decline (as is often seen in the early stages of nova light curves) we derived an approximate $t_{\rm 2}$ time of  2.81$_{-0.9}^{+1.0}$ days (error quoted is statistical only and was obtained using the uncertainties of the fitted gradient and intercept). This is similar to the value given by \citet{Naik}, confirming the very fast classification. Within Figure~\ref{v597pup}, the SMEI light curve is also compared to AAVSO data ({\it V}-band) and data ({\it B}, {\it V}, {\it R} and {\it I}-band) from The STONY BROOK/SMARTS Atlas of (mostly) Southern Novae\footnotemark[2] \citep[hereafter referred to as the Nova Atlas; see][for a full description]{2012PASP..124.1057W} within the same time frame. 

\footnotetext[2]{http://www.astro.sunysb.edu/fwalter/SMARTS/NovaAtlas/}

\begin{figure*}
\centering
\includegraphics[keepaspectratio=true, scale=0.48]{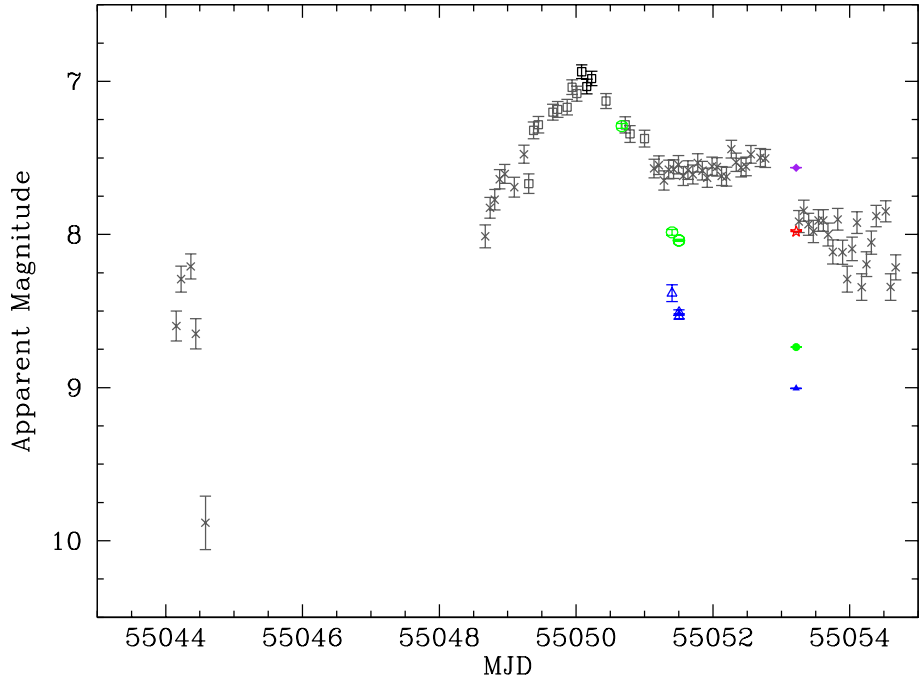}
\caption{Optical light curves of V5583~Sgr. SMEI data with a lower {\it r} than optimal have been included for completeness. Nova Atlas data which coincide with the SMEI light curve have been included along with data from the AAVSO (see Figure~\ref{key} for data legend).}
\label{v5583}
\end{figure*}

\begin{figure*}
\centering
\includegraphics[keepaspectratio=true, scale=0.48]{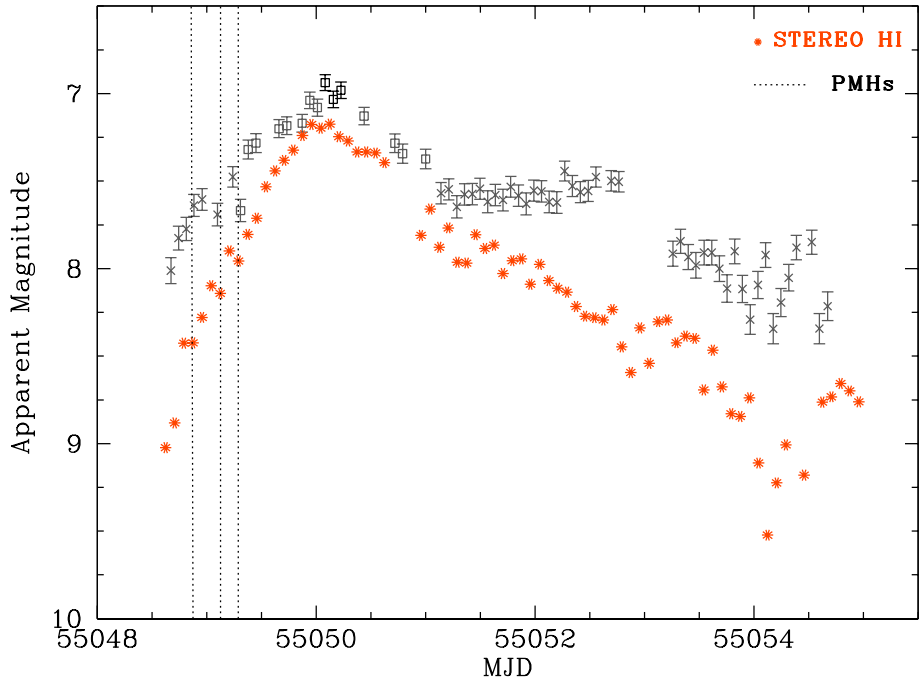}
\caption{Optical light curves of V5583~Sgr. SMEI data with a lower {\it r} than optimal have been included for completeness. The STEREO HI data is given by orange points, the uncertainty of which is 0.06 magnitudes above 10$^{th}$ magnitude. The dotted black lines indicate PMHs suggested in STEREO data \citep{2014MNRAS.438.3483H}, the second of which coincides with that suggested by the SMEI light curve around MJD 55048.98.}
\label{v55832}
\end{figure*}

\begin{figure*}
\centering
\includegraphics[keepaspectratio=true, scale=0.48]{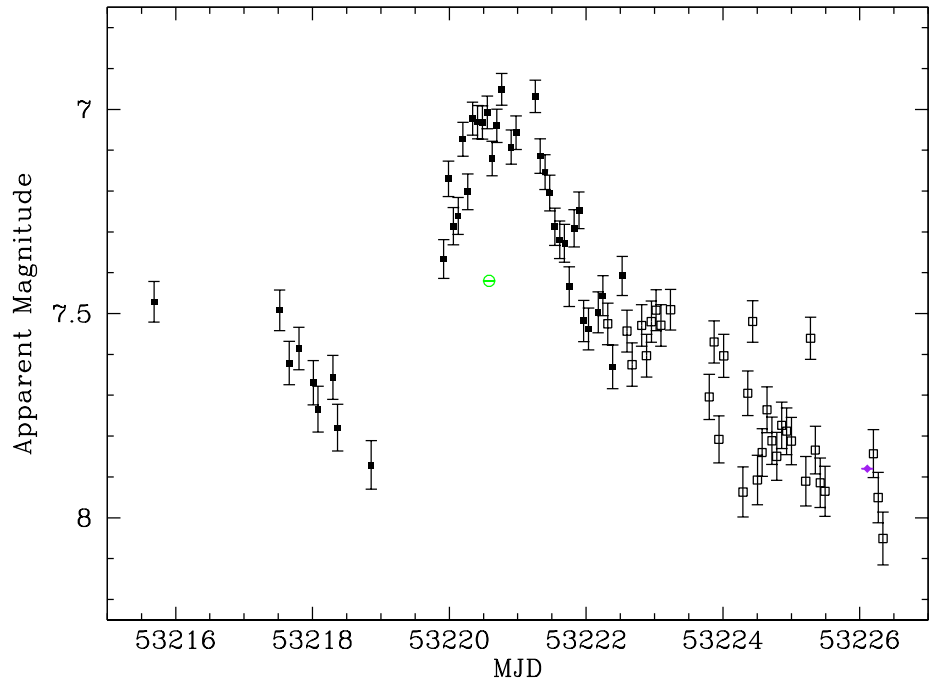}
\caption{Optical light curves of nova V1187~Sco. Nova Atlas data which coincide with the SMEI light curve have been included along with data from the AAVSO (see Figure~\ref{key} for data legend). It is likely that the SMEI data presented here are contaminated by light from neighboring bright stars, and as such should be treated with caution.}
\label{v1187sco}
\end{figure*}

\begin{figure*}
\centering
\includegraphics[keepaspectratio=true, scale=0.48]{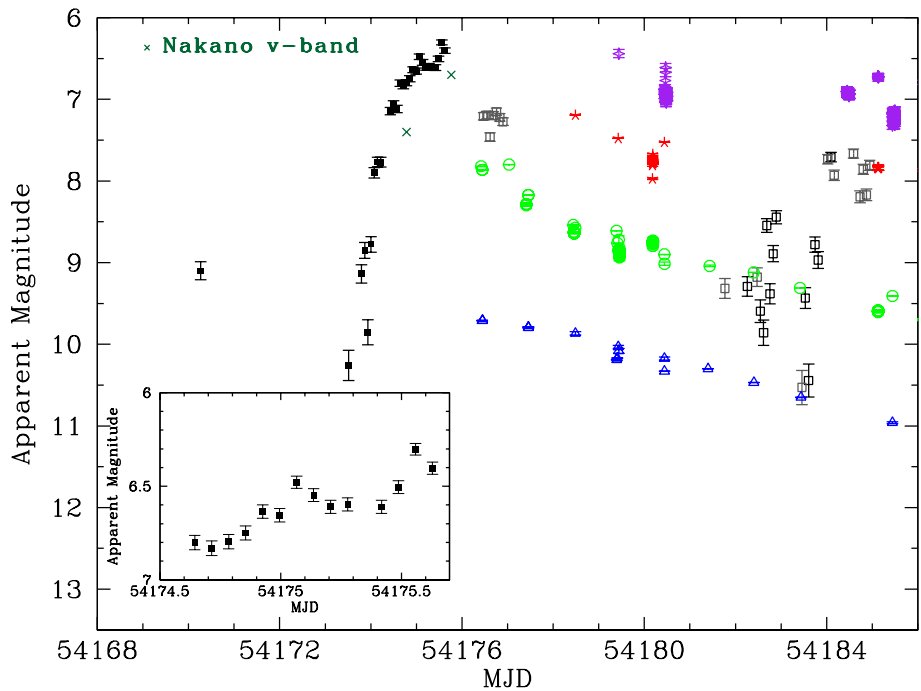}
\caption{Optical light curves of V2467~Cyg indicating the peak of the nova and its initial decline. SMEI data with a lower {\it r} than optimal have been included for completeness. AAVSO data which coincide with the SMEI light curve have been included (see Figure~\ref{key} for data legend). The two green crosses represent data from \citet{Nakano2467}. The inset shows the final rise, peak, and initial decline of the nova in the SMEI data only.}
\label{v2467cyg}
\end{figure*}

\subsection{V5583 Sagittarii}
\label{subsub:v5583Sgr}

Nova V5583~Sgr ($\alpha = 18^{h}07^{m}07^{s}\!\!_{.}67, \;\delta = -33^{\circ}46^{\prime}33^{\prime\prime}\!\!_{.}$9; J2000) was observed by both SMEI and the Solar TErrestrial RElations Observatory (STEREO) Heliospheric Imager \citep{2009SoPh..254..387E}. A review of STEREO observations for this object can be found in \citet{2014MNRAS.438.3483H}.

V5583~Sgr is a very fast CN which was discovered by \citet{Nishiyama} at a magnitude of 7.7 on 2009 August 6.49 UT (MJD 55049.49), it then rose to maximum on 2009 August 7.57 UT (MJD 55050.57) at m$_{V}$=7.43. An approximate $t_{\rm 2}$ time of 5 days is given by \citet{2011ApJS..197...31S}.

The nova was detected by SMEI on 2009 August 1.16 UT (MJD 55044.16) at $m_{\mathrm{SMEI}} = 8.60 \pm 0.10$. There is then a gap in the data due to a combination of low {\it r} and {\it npsf} values lasting until August 5.67 UT (MJD 5548.67) after which it then rises (Figure~\ref{v5583}) to a peak magnitude of $m_{\mathrm{SMEI}} = 6.94 \pm 0.05$ on 2009 August 7.08 $\pm$ 0.04 UT (MJD 55050.08). The timing of this peak coincided nicely with that presented in \citet[][the STEREO peak is given as $m_{\rm HI}$ = 7.18 $\pm$ 0.06 on 2009 August 7.04 $\pm$ 0.17 UT, MJD 55050.04]{2014MNRAS.438.3483H}, however the timing of both the SMEI and STEREO peaks are earlier than that given by \citet{Nishiyama}, suggesting that the maximum was previously missed. 

A PMH may be present in the SMEI data around 2009 August 5.98 UT (MJD 55048.98) at an average $m_{\mathrm{SMEI}} = 7.64$ $\pm$ 0.03 (defined as the mean magnitude over the duration of the halt; quoted error is the rms scatter). Unfortunately, the feature is derived from data with a much lower {\it r} than optimal ({\it r} values are 0.23, 0.26, and 0.26 respectively) and is therefore unreliable. However, it should be noted that \citet{2014MNRAS.438.3483H} also find evidence of PMHs around this time. \citet{2014MNRAS.438.3483H} define a PMH as a decrease in the brightness during the rise phase of the nova eruption and detect three possible events, the first on MJD 55048.88 $\pm$ 0.08 at $m_{\rm HI}$=8.38 $\pm$ 0.09, the second on MJD 55049.13 $\pm$ 0.08 at $m_{\rm HI}$=8.05 $\pm$ 0.10 (which when taking into consideration timing definitions, coincides with the SMEI PMH), and the third on MJD 55049.29 $\pm$ 0.08 at  $m_{\rm HI}$=7.89 $\pm$ 0.08 (see Figure~\ref{v55832}). This third halt is also seen in the SMEI light curve at an average $m_{\mathrm{SMEI}} = 7.18$. However it is only $\sim$ 0.25 magnitudes below maximum and is not within the range that has generally been proposed for a PMH. 

Approximately one day after maximum the object then seems to enter a ``plateau'' phase which lasts for 1.6 days with an average $m_{\mathrm{SMEI}} = 7.56$, after which it declines. Again it should be noted that these data have a correlation coefficient much less than optimal (0.1~ $\leq$~ {\it r}~ $<$~0.3), and are less reliable. The discrepancy between the SMEI and STEREO light curves especially at late times is clearly evident and due to the reduced quality of the SMEI data ({\it r}~ $<$~ 0.3) throughout the nova eruption. Therefore, the r $<$ 0.3 SMEI data are less reliable and complementary to the STEREO data when obtained simultaneously.

\subsection{V1187 Scorpii}
\label{subsub:v1187Sco}
Nova V1187~Sco ($\alpha = 17^{h}29^{m}18^{s}\!\!_{.}81, \;\delta = -31^{\circ}46^{\prime}$01$^{\prime\prime}\!\!_{.}$5; J2000) was discovered prior to peak at $m_{\rm {V}}$ = 9.9 on 2004 August 2.07 UT (MJD 53219.07) using data from the All Sky Automated Survey (ASAS)-3 patrol \citep{2005AcA....55..275P}. It subsequently rose to a maximum magnitude of $m_{\rm {V}}$ = 7.42 on 2004 August 3.58 UT \citep[MJD 53220.58,][this is the peak of the V band AAVSO light curve given in Figure~\ref{v1187sco}]{Yamaoka04}. The initial decline gave $t_{\rm {2}}$ = 8.7 and $t_{\rm {3}}$ = 15 days \citep{Lynch1187}, classifying V1187 Sco as a very fast nova. Near-IR spectroscopic observations of the object by \citet{Lynch1187} indicated the development of a nova explosion on an O~Ne WD, which did not form dust before entering its nebular phase. The emission lines found within the spectra had complex, double-peaked profiles. Using the H~I double emission lines the nova ejecta were modeled as ring or partial sphere-like emitting region. An extinction of $A_{\rm {V}}$ = 4.68 $\pm$ 0.24 was derived also using O~I lines in combination with the optical spectra.

The SMEI data give the peak brightness of nova V1187~Sco as  $m_{\mathrm{SMEI}} = 6.95 \pm 0.04$ on 2004 August 3.77$_{-0.04}^{+0.07}$ UT (MJD 53220.77). This is 0.47 magnitudes brighter and 0.19 days later than the {\it V}-band peak given in the lower time resolution results of \citet{Yamaoka04}. Within Figure~\ref{v1187sco}, the SMEI light curve is compared to {\it V}-band AAVSO data and {\it I}-band data from the Nova Atlas within the same time frame. 

A substantial portion of the initial nova rise is missing from the SMEI light curve, however some data before peak are presented. As the nova is of the very fast classification one would expect the rise to maximum to also be fast, however within the SMEI data we are finding the opposite effect. This apparent slow rise is most likely due to contamination of the data from several close bright neighboring stars, despite the conduction of simultaneous fitting and the fitting of additional bright stars in the larger surrounding region. The apparent plateau/dip before the final rise to peak may therefore not be a PMH, instead it is probably residual light from a neighboring star. Data points between July 29$^{th}$ to July 31$^{st}$ (MJD 53215-53217) were removed due to large variations in the RA and dec on fitting, suggesting a fit of residuals rather than the nova. However, the gap seen around the 2$^{nd}$ of August (MJD 53219) is due to a SMEI anomaly during which the instrument was forced to shut down and restart\footnotemark[3]. The noisiness of the light curve especially during decline is also likely due to contamination from neighboring bright stars. However, there does appear to be some structured oscillation which may be real.

The SMEI data shown in Figure~\ref{v1187sco} do not cover the magnitude range required for the determination of a $t_{\rm 2}$ time. However, on application of a liner fit to the initial decline of the nova using all data where {\it r} is greater than 0.3 a $t_{\rm 2}$ of 10.10$_{-0.39}^{+0.43}$ days was derived (error quoted is statistical only and obtained using the uncertainties of the fitted gradient and intercept.). This is slightly larger than the result presented in \citet{Lynch1187}, but agrees (within errors) with the very fast nova classification. 

\footnotetext[3]{A SMEI anomaly means that satellite underwent an impromptu manipulation causing the system to go offline.}

\subsection{V2467 Cygni}
\label{subub:v2467Cyg}
Nova V2467~Cyg ($\alpha$ = $20^{h}28^{m}12^{s}\!\!_{.}47,\;$ $\delta$ = $+41^{\circ}48^{\prime}36^{\prime\prime}\!\!_{.}$4; J2000) was discovered by A. Tago at $m_{\rm {V}}$ = 7.4 \citep{Nakano2467} on 2007 March 15.79 UT (MJD 54174.79). It then rose to a peak magnitude of 6.7 on 2007 March 16.77 UT (MJD 54175.77) and declined with a $t_{\rm 2}$ = 7.3 \citep{2009AAS...21349125L}. An early spectrum of the nova obtained on 2007 March 16.8 UT (MJD 54175.8) indicated an expansion velocity of $\sim$ 1200 km s$^{-1}$ and an Fe~II - type nova classification \citep{Munari2467}. 

The transition phase of the object was seen to start in April of 2007 after fading approximately 4 magnitudes. Within this phase, six quasi-periodic oscillations were observed with periods from 19 to 25 days and amplitudes of $\approx$ 0.7 magnitudes. \citet{Swierczynski} proposed that the period found within the optical light curve, and changes found in the subsequent X-ray detections, could only be explained if the system were an IP.

The SMEI light curve for nova V2467~Cyg can be seen in Figure~\ref{v2467cyg}. The data show a gap between the first detectable point and the initial rise of the nova. Examination of data within this period suggests that the SMEI pipeline struggled to find a point source due to a combination of noise from nearby objects and the overall faintness of the nova at this time. The initial point seen in Figure~\ref{v2467cyg} is only just detectable above this background noise. The rise of the nova light curve starts on 2007 March 14.52 UT (MJD 54173.52) and is very steep, rising 2.8 magnitudes in just under 2 days. Within this rise to peak no PMH is found, possibly due to a lack of data around $m_{\rm SMEI}$ = 7.5. The peak magnitude of the nova is given as  $m_{\mathrm{SMEI}} = 6.30 \pm 0.03$ on 2007 March 16.56 $\pm 0.04$ UT (MJD 54175.56), this is 0.40 magnitudes brighter and 0.21 days earlier than the {\it V}-band peak given in \citet{Nakano2467} and suggests that it was not covered within their data. Once again the large scatter of the SMEI data on decline prevents a direct measurement of t$_{2}$. Assuming that any scatter in the data is due to noise and is not inherent to the intrinsic behavior of the nova a linear fit of the initial decline yields an approximate $t_{\rm 2}$ time of 5.68$_{-0.53}^{+0.61}$ days, classifying the nova as very fast.  This is a sharper decline than that observed by \citet{2009AAS...21349125L}. During this decay there are several gaps in the SMEI data. The first takes place during March 16$^{th}$ (MJD 54175) and is due to a masking issue which lasts for less than a day, the second occurs on March 18$^{th}$ (MJD 54177) and is due to many factors including engineering work, pre-annealing calibration, hot annealing, and an instrumental anomaly. This gap lasted for several days.

\begin{figure*}
\centering
\includegraphics[keepaspectratio=true, scale=0.5]{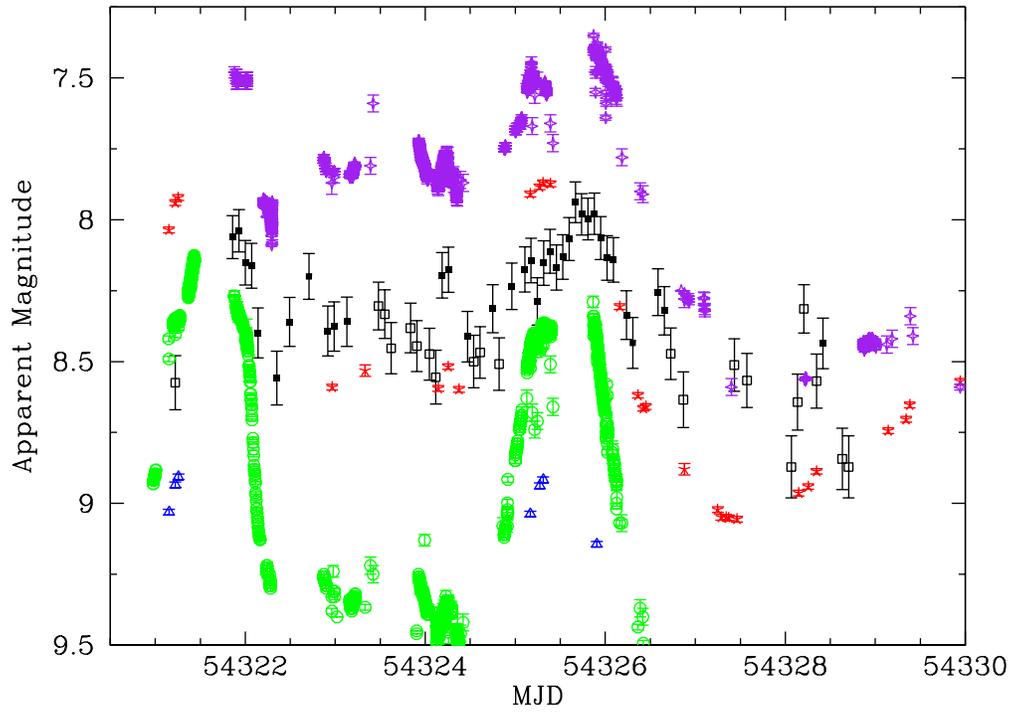}
\caption{Optical light curves of V458~Vul indicating the multiple peaks of the nova. AAVSO  data are displayed for comparison (see Figure~\ref{key} for data legend). There is a large color difference between blue and red filters.}
\label{v458vul}
\end{figure*}

\begin{figure*}
\centering
\includegraphics[keepaspectratio=true, scale=0.5]{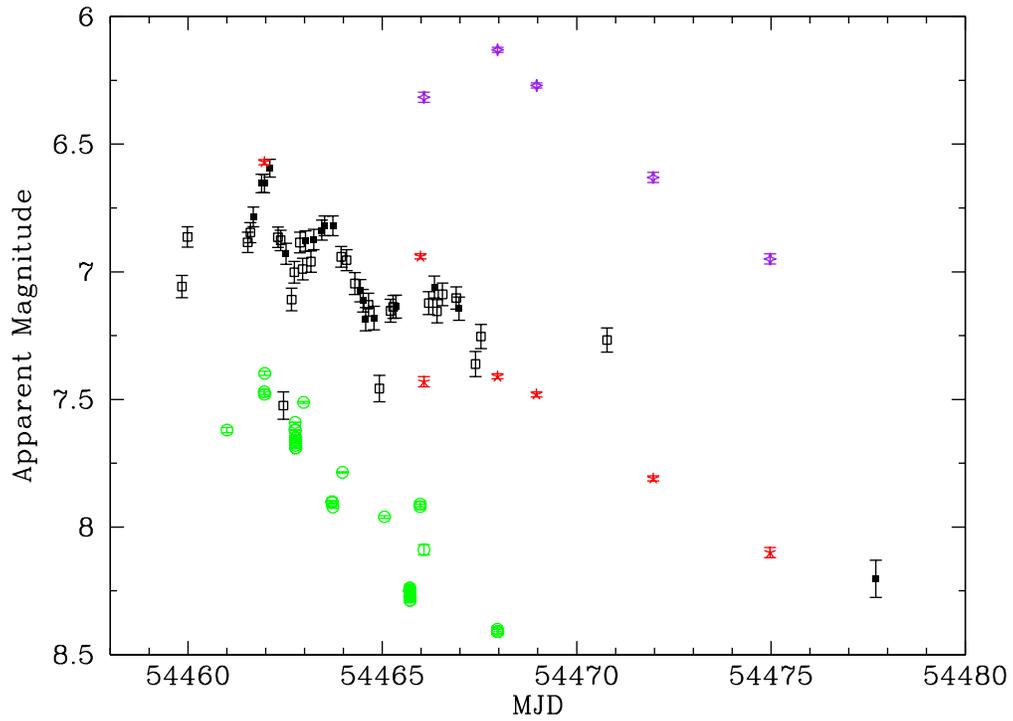}
\caption{Optical light curves of V459~Vul. The SMEI light curve indicates oscillations during the decline from maximum. Data from the AAVSO are displayed for comparison (see Figure~\ref{key} for data legend).}
\label{v459vul}
\end{figure*}

\subsection{V458 Vulpeculae}
\label{subsub:v458Vul}
Nova V458~Vul ($\alpha$ = $19^{h}54^{m}24^{s}\!\!_{.}61$, $\delta$ = $+20^{\circ}52^{\prime}52^{\prime\prime}\!\!_{.}$6; J2000) was discovered in TN-outburst by \citet{2007IAUC.8861....2N} at an apparent magnitude of 9.5 on 2007 August 8.54 UT (MJD 54320.54), reaching its peak visual magnitude soon after at m$_{\rm V}$ = 8.1. The nova's $t_{\rm 3}$ time is given as 21 days and it is therefore classified as a fast nova \citep{2008ApJ...688L..21W}. The decline of the object was disrupted by two re-brightenings, the first occurring 20 days after maximum. Spectra taken after the eruption revealed the nova to be of the hybrid spectral class \citep{2008Ap&SS.315...79P}, with an early Fe~II - type spectra and evolving to He/N. V458 Vul is an interesting nova as it occurred within a planetary nebula and as such light echoes of the burst within the surrounding material are seen \citep{2008ApJ...688L..21W}. 

The SMEI light curve and corresponding AAVSO data of this nova are displayed in Figure~\ref{v458vul}. Unfortunately SMEI covers only eight days around the nova peak, as on other dates the nova is too faint for detection. What is most striking about this light curve is that there are multiple flaring events during the initial decline \citep[these may be akin to those seen in U Sco,][]{2011ApJ...742..113S}. There are two bright peaks separated by a fainter third, the magnitude and times of each peak are given as $m_{\rm SMEI} = 8.04 \pm 0.07$ on 2007 August 9.93 $\pm$ 0.04 UT (MJD 54321.93), $m_{\rm SMEI} = 8.30 \pm 0.08$ on 2007 August 11.48$_{-0.18}^{+0.04}$ UT (MJD 54323.48), and  $m_{\rm SMEI} = 7.94 \pm 0.07$  on 2007 August 13.66 $\pm$ 0.04 UT (MJD 54325.66). On comparison the peak of the {\it R}-band light curve occurs on 2007 August 13.31 UT (MJD 54325.31) at m$_{\rm R}$ = 7.87 $\pm$ 0.01, with the peak of the {\it I} on 2007 August 13.87 UT (MJD 54325.87) at  m$_{\rm I}$ = 7.35  $\pm$ 0.01.

\subsection{V459 Vulpeculae}
\label{v459Vul}
Nova V459 Vul ($\alpha = 19^{h}48^{m}08^{s}\!\!_{.}87, \;\delta = +21^{\circ}15^{\prime}26^{\prime\prime}\!\!_{.}$8; J2000) was discovered independently by Hiroshi Kaneda and Akihiko Tago \citep{Nakano1, Nakano2} at an average unfiltered magnitude of 8.7 on 2007 December 25.35 UT (MJD 54459.35) and 2007 December 26.38 UT (MJD 54460.38) respectively. Spectroscopic observation of the object by \citet{Yamaokav459vul} revealed the presence of several Fe~II multiplets making V459~Vul an Fe~II - type nova. A candidate progenitor with $m_{\rm {r}} \sim$ 20 was identified in the red POSS-II plates, but no IR counterpart was found within 2MASS. The progenitor magnitude found suggested a TN-outburst amplitude of $\sim$ 12.5 magnitudes in B \citep{2008IBVS.5822....1H}. The maximum magnitude of V459~Vul was $m_{\rm {V}}$ = 7.58 on 2007 December 27.25 UT (MJD 54461.25) with $t_{\rm {2}}$ and $t_{\rm {3}}$ times given as 18 $\pm$ 2 days and 30 $\pm$ 2 days respectively \citep{Poggiani}, making this a fast nova. Using photometric data, \citet{Poggiani} went on to obtain an extinction of $A_{\rm {V}}$ = 2.75 $\pm$ 0.38, an absolute magnitude range between -8.7 and -7.7, and a WD mass in the range of 0.9 - 1.1 M$_\odot$.

The SMEI light curve is displayed in Figure~\ref{v459vul} and unfortunately the initial rise of the nova appears not to be caught by the instrument. Several points between December 25.85 UT and 27.54 UT (MJD 54459.85 - 54461.54) have also been omitted due to poor {\it r} values and a large scatter in the RA and dec of the object early on, as such any possible PMH has been missed. The peak magnitude of the light curve is found to be $m_{\mathrm{SMEI}} = 6.59 \pm 0.04$ on 2007 December 28.11$_{-0.07}^{+0.11}$ UT (MJD 54462.11). Based on a linear extrapolation between the peak and last data point of the SMEI light curve the estimated $t_{\rm 2}$ time is given as $\sim$ 19.4 days. The SMEI peak is 0.99 magnitudes brighter and 0.86 days later than {\it V}-band measurements given in \citet{Poggiani}, and the estimated $t_{\rm 2}$ value is slightly larger, these deviations may be due to the difference in band-pass. The peak AAVSO {\it V}-band magnitude however, disagrees with that of \citet{Poggiani} giving m$_{\rm V}$ = 7.39 $\pm$ 0.01 on 2007 December 27.98 (MJD 54461.98) reducing previous discrepancies.  On comparison the peak of the {\it R}-band AAVSO data is much closer to the SMEI peak as it occurs on 2007 December 27.97 UT (MJD 54461.97) at m$_{R}$=6.57 $\pm$ 0.01. The SMEI light curve also indicates that there may be several oscillations in the decline with amplitudes of order of a few tenths of a magnitude. These oscillations are not evident in the AAVSO data and illustrate how high cadence observations can reveal new features.

\begin{figure}
\centering
\includegraphics[keepaspectratio=true, scale=0.32]{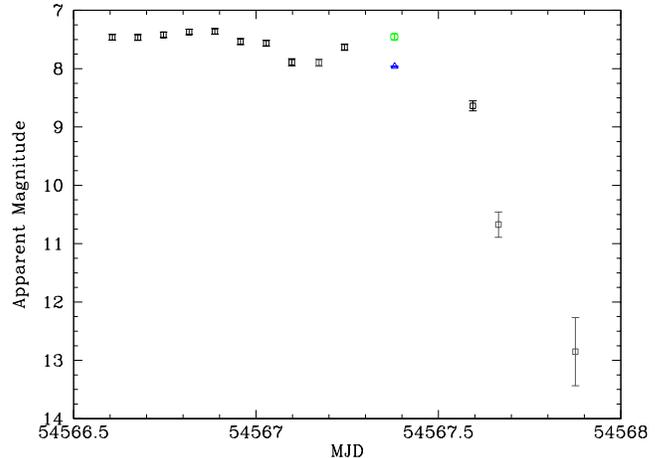}
\caption{Optical light curves of V2491~Cyg. SMEI data with a lower {\it r} than optimal have been included for completeness. Data from the AAVSO have been plotted for comparison (see Figure~\ref{key} for data legend).}
\label{v2491cyg}
\end{figure}

\begin{figure*}
\centering
\includegraphics[keepaspectratio=true, scale=0.6]{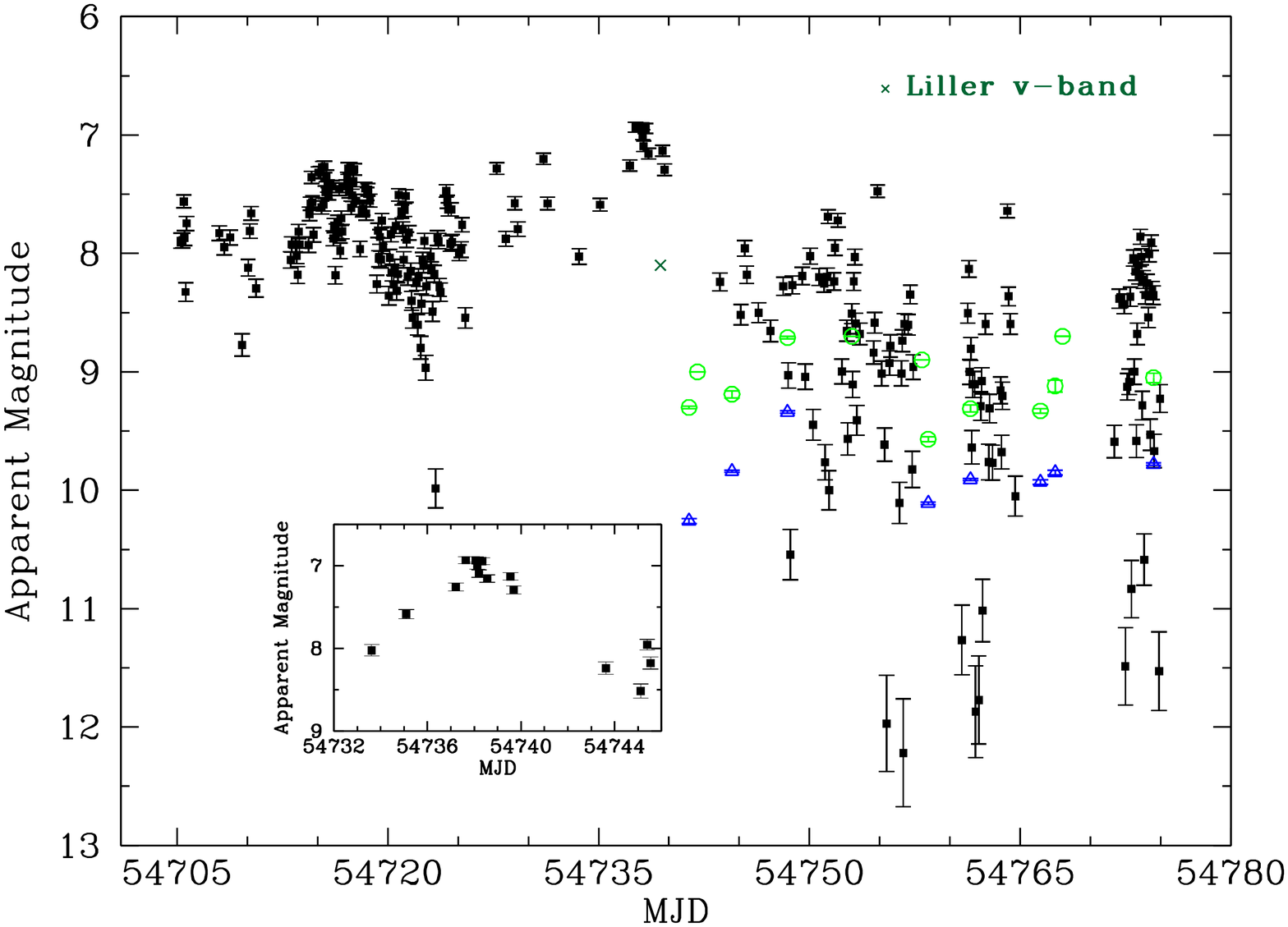}
\caption{Optical light curves of QY~Mus. The SMEI data are very noisy due to contamination from neighboring bright stars}. AAVSO data are given for comparison (see Figure~\ref{key} for data legend). The green cross represent the peak unfiltered magnitude given by \citet{Liller}. The inset represents data from around the peak of the SMEI light curve only. {\bf These data at maximum are to be treated with caution as they are contaminated by light from neighboring bright stars.}
\label{qymus}
\end{figure*}

\subsection{V2491 Cygni}
\label{subsub:v2491cyg}
The discovery of Nova V2491~Cyg ($\alpha$ = $19^{h}43^{m}01^{s}\!\!_{.}96,\;$ $\delta$ = $+32^{\circ}19^{\prime}13^{\prime\prime}\!\!_{.}$8; J2000) was presented in \citet{2008IAUC.8934....1N}. The nova reached maximum $m_{\rm V}$ = 7.45 $\pm$ 0.05 on 2008 April 11.37 $\pm$ 0.01 UT \citep[MJD 54567.37, this is the peak of the AAVSO {\it V}-band data in Figure~\ref{v2491cyg},][]{2011NewA...16..209M}. Spectra indicated that it belongs to the He/N class of novae \citep{2008CBET.1379....1H}. V2491~Cyg had a rapid optical decline with a $t_{\rm 2}$ = 4.8 days and as such is classified as a very fast nova \citep{2011NewA...16..209M, 2011A&A...530A..70D}. Approximately 15 days after eruption a secondary maximum was observed with $m_{\rm {V}}$ = 9.49 $\pm$ 0.03. Based on the spectra of the nova many authors \citep[e.g.][]{2008ATel.1485....1T, 2010MNRAS.401..121P} believe that the object is in fact a RN. Using the interstellar Na~I line, a reddening of $E_{\rm {B-V}}$ = 0.23 $\pm$ 0.01 was derived.

Unfortunately SMEI was only able to detect the nova at a few points around the nova peak and its initial decline. These data are presented in Figure~\ref{v2491cyg}, in which we have allowed data with {\it r} $<$ 0.4 for completeness. The maximum magnitude is given as $m_{\mathrm{SMEI}} = 7.36 \pm 0.05$ on 2008 April 10.89 $\pm$ 0.04 UT (MJD 54566.89), slightly earlier and brighter than the {\it V}-band peak given in \citet{2011NewA...16..209M}, but matching quite well the unfiltered magnitude of 7.7 on 2008 April 10.8 UT (MJD 54566.8) presented in \citet{2008IAUC.8934....1N}. Due to a lack of data we were unable to determine a  $t_{\rm 2}$ time.

\subsection{QY Muscae} 
\label{subsub:qymus}
QY~Mus ($\alpha = 13^{h}16^{m}36^{s}\!\!_{.}44, \;\delta = -67^{\circ}36^{\prime}47^{\prime\prime}\!\!_{.}$8; J2000) was discovered by \citet{Liller}, at an unfiltered magnitude of 8.6 on 2008 September 28.998 UT (MJD 54737.998). The nova then reached a peak unfiltered magnitude of 8.1 on 2008 September 30.397 UT (MJD 54739.40, green cross within Figure~\ref{qymus}). \citet{2011ApJS..197...31S} give the $t_{\rm 2}$ time of this object as approximately 60 days, making this a moderately fast nova.

The SMEI light curve (see Figure~\ref{qymus}) indicates that the nova reached a peak magnitude of $m_{\mathrm{SMEI}} = 6.93 \pm 0.04$ on 2008 September 28.63 $\pm$ 0.2 UT (MJD 54737.63). This magnitude is significantly brighter than both the {\it V}-band AAVSO peak and unfiltered peak given in \citet{Liller}. The variance between the peak magnitudes may be due to differences in the band-passes of the instruments. However, upon an examination of the location of the nova within the SMEI sky-maps, the discrepancy is most likely caused by neighboring bright stars (again, simultaneous fitting of these objects was conducted along with the fitting of additional bright stars in the larger surrounding region). The decay of the SMEI light curve is quite consistent with that from AAVSO data. However the scatter in the SMEI light curve is evident and is due to problems in fitting the source as it approaches the limiting SMEI magnitude, and contamination from the surrounding bright stars. We are unable to give an estimate of the $t_{\rm 2}$ value of the nova due to the large scatter in the light curve at this time. 

\subsection{V5580 Sagittarii}
\label{subsub:v5580Sco}
Nova V5580~Sgr ($\alpha = 18^{h}22^{m}01^{s}\!\!_{.}39, \;\delta = -28^{\circ}02^{\prime}39^{\prime\prime}\!\!_{.}$8; J2000) was discovered by \citet{Liller5580} at approximately 8$^{th}$ magnitude using an ``orange" filter on 2008 November 29.04 UT (MJD 54799.04). The variable was also present on November 23.037 UT (MJD 54793.037) at a magnitude of approximately 10.3, but was not visible (mag $>$ 11.0) on November 20.035 (MJD 54790.035). The nova rose to peak on 2008 November 29.999 UT (MJD 54799.999) with a magnitude of 7.8 \citep{Liller5580}.

Although the nova is detected by SMEI, only a few reliable points were obtained due to the passage of Venus and its poor subtraction. For this reason we have decided not to present the light curve. The peak magnitude found for the object is at $m_{\rm SMEI} = 7.01 \pm 0.04$ on 2008 November 30.85 UT (MJD 54800.85). This is slightly later and brighter than values mentioned within \citet{Liller5580}. Unfortunately no $t_{\rm 2}$ time can be derived from the SMEI light curve, and one can not be found in the literature.

\section{Discussion and Conclusion}
\label{disscussion_conculsion}
We present here SMEI archival data for nine known Galactic novae, V1187~Sco, V2467~Cyg, V458~Vul, V597~Pup, V459~Vul, V2491~Cyg, QY~Mus, V5580 Sgr, and V5583~Sgr \citep[which are faint with respect to the novae presented in Paper I: RS~Oph, V1280~Sco, V598~Pup, and KT Eri, and comparable to T~Pyxidis in][]{2012IAUS..285...91H, 2014AJ....147..107S}. Light curves for eight of the eruptions have been displayed, and for all nine we have determined basic observational properties, including the time of peak, maximum magnitude, and where possible the $t_{\rm 2}$ time ({see Table~\ref{table1}).  Our work indicates that although these novae possess magnitudes which are at or below the optimal detection limits of SMEI, we are still able to produce light curves which in many cases contain more data at and around the initial rise, peak, and decline than are found within other variable star catalogs such as the AAVSO or the Nova Atlas (the limits here may well be due to sampling and distance).

The majority of novae examined within this paper are of the very fast (5 out of 9) or fast (2 of 9) speed classes rather than the moderately fast, slow, or very slow type. This is not surprising as novae which decline more rapidly are more intrinsically luminous and as such more likely to possess peak magnitudes above the SMEI detection threshold (distance and reddening factors may also play a part here as well). QY~Mus, however does belong to the moderately fast speed class, which in turn implies a slow photometric evolution (the slowest nova examined in both this paper and Paper I), and as such has produced one of the noisiest light curves due to the faintness of the event (and star crowding issues). 

Whenever possible our SMEI light curves have been compared to multi-filter data from the AAVSO and Nova Atlas. In the majority of cases the SMEI data have been seen to match the {\it R}-band (specifically $R_{\rm AAVSO}$) much better than those of any other filter (unsurprising given that the peak throughput of the instrument is 700nm). However, for V1187~Sco the SMEI magnitude is best matched with the {\it I}-band Nova Atlas data. This discrepancy between favored band-passes could be caused by contamination from neighboring stars, as well as different filters being used in both the Nova Atlas and AAVSO catalogs, which come from several different telescopes, and as such vary in peak throughput. It could also be due to the intrinsic spectral energy distribution of the nova, and/or the column of extinction towards the system. An additional point to note is that V1187 Sco is a fast ONe type nova, and so would have had much stronger line emission early on in the eruption relative to slower, more continuum dominated novae. The strong H$\alpha$ emission would effect the {\it R} and {\it I} filters and may well  be a more likely explanation for the difference in which one is favored.

Four of the novae examined (V1187~Sco, V458~Vul, V459~Vul, and V5580~Sgr) possess peak SMEI magnitudes which are brighter than their recorded {\it V}-band maxima, and at slightly later times. Again this may simply be due to the different filters being considered here, which is supported by the fact that two of the novae possess peaks which are closer in time and magnitude to the given {\it R}-band maxima. On the other hand four novae (QY~Mus, V2467~Cyg, V2491~Cyg, V5583~Sgr) have peak SMEI magnitudes which are earlier and brighter than the values recorded in the literature. For QY~Mus we believe that the value of the SMEI maximum is inaccurate due to severe contamination from neighboring sources. However, for V2467~Cyg, V2491~Cyg, and V5583~Sgr, although the difference in magnitude may be due to the difference in band-pass, the earlier occurrence could suggest that we obtained a more precise time of peak due to better sampled light curves. 

Looking at specific features within the SMEI nova light curves, five of the objects examined possess final rise data (V1187~Sco, V2467~Cyg, V597~Pup, V459~Vul, V5583~Sgr), which for the majority of cases (V5583~Sgr excluded) were not previously observed. Within the final rise of both V597~Pup and V5583~Sgr the presence of a PMH is suggested. Unfortunately the correlation coefficient ({\it r}) of the data for the PMHs is in both cases less than optimal (the first two points of the V597~Pup halt have an {\it r} value of  0.39 and 0.35; the V5583~Sgr halt has {\it r} values of 0.23, 0.26, and 0.26 respectively) and so their reality is questionable. However, in the case of V5583~Sgr work by \citet{2014MNRAS.438.3483H} would suggest that the SMEI PMH is real as they too find this halt using STEREO data. It should also be noted that the duration of each halt observed is consistent with the speed class of the nova, and occurs within the $\Delta m_{\rm SMEI}$ (0.63-1.74 magnitudes) and $\Delta t$  (0.49 - 2.19 days) ranges found in Paper I.

Throughout this work and Paper I, we have used the traditional definition of a PMH which occurs 1-2 mags below peak (see e.g. McLaughlin).  However, we should note that until there is a strong physical basis for the PMH any such definition is relatively arbitrary, and, for example, maybe depend on parameters such as the WD mass or accretion rate.

With this in mind we must re-evaluate both nova V2467~Cyg and V5583~Sgr. Examining the rise of nova V2467~Cyg there may be evidence for a PMH on 2007 March 16.07 UT (MJD 54175.07) at an average $m_{\rm SMEI} = 6.56 \pm 0.05$ which lasts $\sim0.14$ days (see inset of Figure~\ref{v2467cyg}). This PMH has a $\Delta m_{\rm SMEI}$ of 0.26 magnitudes, much smaller than that previously observed, and a $\Delta t$ of 0.49 days. The {\bf r} values associated with this PMH data are also all above 0.6 making it a very real feature of the light-curve. In Section~\ref{subsub:v5583Sgr} nova V5583 Sgr was found to have several PMH like features, and although one seemed to fit our `standard' definition of a PMH it consisted of data with lower {\bf r} values than optimal. The third feature seen however, has data with {\bf r} $>$ 0.35 (again less than optimal, but better than the earlier features) and occurs on 2009 August 6.75 UT (MJD 55049.75) at $m_{\rm SMEI} = 7.18 \pm 0.01$ and lasts for 0.21 days. The PMH has an $\Delta m_{\rm SMEI}$ of 0.25 magnitudes and a $\Delta t$ of 0.33 days, both of which are much smaller than that previously observed.

During the initial decline from maximum both V1187~Sco and V459~Vul seem to display oscillations within their light curves. This rapid observed variability can only be appreciated and realized with a high-cadence instrument such as SMEI. Moreover, the unusual nova V458~Vul has several multi-flaring events shortly after eruption for which SMEI has again provided a highly detailed light curve. This light curve closely matches that observed in the AAVSO {\it R}-band. 

With its closure in September 2011 the SMEI archive contains 8.5 years of all-sky high cadence (102 minutes) data which is now in a static state. Both Paper I and this present work clearly indicate how important it is to examine the data of all-sky facilities such as SMEI with regards to transient events. As such we will continue to investigate this archive for other transient and variable stars, as well as for additional novae which may have been missed during their eruptions.

\section{Acknowledgements}
R. Hounsell acknowledges support from the Space Telescope Science Institute and the University of Illinois Urbana Champaign.
The USAF/NASA SMEI is a joint project of the University of California San Diego, Boston College, the University of Birmingham (UK), and the Air Force Research Laboratory.
P. P. Hick, A. Buffington, B. V. Jackson, and J. M. Clover acknowledge support from NSF grant ATM-0852246 and NASA grant NNX08AJ11G. A. W. Shafter acknowledges support from NSF grant AST-1009566. 
S. Starrfield acknowledges partial support from NASA and NSF grants to ASU.
We would like to thank Dr. Zach Cano (Centre for Astrophysics and Cosmology at the University of Iceland) for his thoughtful ideas and help during this project. Finally we would like to thank the referee for their helpful comments and guidance.


\clearpage

\appendix
This appendix contains the MJD, SMEI Mag, and SMEI Error for each nova examined within this paper and Paper~I. In addition {\bf r} is given for novae presented within this paper only.

\begin{table}
\centering
\scriptsize
\caption{SMEI light curve data for Nova V1187 Sco}
\hrule
\begin{minipage}[t]{0.4\linewidth}
\begin{tabular}[t]{lccc}
Date-Obs MJD & SMEI Mag & SMEI Error & {\bf r} \\
\hline
53215.68 & 7.47 & 0.05 & 0.73 \\
53217.52 & 7.49 & 0.05 & 0.72 \\
53217.66 & 7.62 & 0.05 & 0.70 \\
53217.80 & 7.59 & 0.05 & 0.71 \\
53218.01 & 7.67 & 0.05 & 0.71 \\
53218.08 & 7.73 & 0.06 & 0.69 \\
53218.30 & 7.66 & 0.05 & 0.70 \\
53218.37 & 7.78 & 0.06 & 0.76 \\
53218.86 & 7.87 & 0.06 & 0.74 \\
53219.92 & 7.37 & 0.05 & 0.63 \\
53219.99 & 7.17 & 0.04 & 0.64 \\
53220.06 & 7.29 & 0.05 & 0.64 \\
53220.13 & 7.26 & 0.05 & 0.63 \\
53220.20 & 7.07 & 0.04 & 0.66 \\
53220.27 & 7.20 & 0.04 & 0.61 \\
53220.34 & 7.02 & 0.04 & 0.65 \\
53220.41 & 7.03 & 0.04 & 0.63 \\
53220.48 & 7.03 & 0.04 & 0.64 \\
53220.55 & 7.01 & 0.04 & 0.64 \\
53220.62 & 7.12 & 0.04 & 0.64 \\
53220.70 & 7.04 & 0.04 & 0.62 \\
53220.77 & 6.95 & 0.04 & 0.63 \\
53220.91 & 7.09 & 0.04 & 0.57 \\
53220.98 & 7.06 & 0.04 & 0.60 \\
53221.26 & 6.97 & 0.04 & 0.61 \\
53221.33 & 7.11 & 0.04 & 0.58 \\
53221.40 & 7.15 & 0.04 & 0.58 \\
53221.47 & 7.20 & 0.04 & 0.54 \\
53221.54 & 7.29 & 0.05 & 0.57 \\
53221.61 & 7.32 & 0.05 & 0.57 \\
53221.68 & 7.33 & 0.05 & 0.57 \\
53221.75 & 7.43 & 0.05 & 0.55 \\
53221.82 & 7.29 & 0.05 & 0.55 \\
53221.89 & 7.25 & 0.04 & 0.56 \\
53221.96 & 7.52 & 0.05 & 0.53 \\
53222.04 & 7.54 & 0.05 & 0.52 \\
53222.18 & 7.50 & 0.05 & 0.51 \\
53222.25 & 7.46 & 0.05 & 0.52 \\
53222.32 & 7.53 & 0.05 & 0.50 \\
53222.39 & 7.63 & 0.05 & 0.50 \\
53222.53 & 7.41 & 0.05 & 0.51 \\
53222.60 & 7.54 & 0.05 & 0.49 \\
53222.67 & 7.63 & 0.05 & 0.47 \\
53222.81 & 7.53 & 0.05 & 0.50 \\
53222.88 & 7.60 & 0.05 & 0.48 \\
    \end{tabular}
\end{minipage}
\begin{minipage}[t]{0.4\linewidth}
	\begin{tabular}[t]{lccc}
Date-Obs MJD & SMEI Mag & SMEI Error & {\bf r} \\
\hline
53222.95 & 7.52 & 0.05 & 0.49 \\
53223.02 & 7.49 & 0.05 & 0.49 \\
53223.09 & 7.53 & 0.05 & 0.49 \\
53223.23 & 7.49 & 0.05 & 0.47 \\
53223.80 & 7.70 & 0.06 & 0.46 \\
53223.87 & 7.57 & 0.05 & 0.45 \\
53223.94 & 7.81 & 0.06 & 0.43 \\
53224.01 & 7.60 & 0.05 & 0.44 \\
53224.29 & 7.94 & 0.06 & 0.40 \\
53224.36 & 7.70 & 0.05 & 0.44 \\
53224.43 & 7.52 & 0.05 & 0.44 \\
53224.50 & 7.91 & 0.06 & 0.41 \\
53224.57 & 7.84 & 0.06 & 0.43 \\
53224.64 & 7.74 & 0.06 & 0.42 \\
53224.71 & 7.81 & 0.06 & 0.41 \\
53224.79 & 7.85 & 0.06 & 0.40 \\
53224.86 & 7.77 & 0.06 & 0.42 \\
53224.93 & 7.79 & 0.06 & 0.41 \\
53225.00 & 7.81 & 0.06 & 0.42 \\
53225.21 & 7.91 & 0.06 & 0.42 \\
53225.28 & 7.56 & 0.05 & 0.48 \\
53225.35 & 7.83 & 0.06 & 0.44 \\
53225.42 & 7.91 & 0.06 & 0.42 \\
53225.49 & 7.94 & 0.06 & 0.42 \\
53226.20 & 7.84 & 0.06 & 0.44 \\
53226.27 & 7.95 & 0.06 & 0.43 \\
53226.34 & 8.05 & 0.06 & 0.42 \\
53226.48 & 8.30 & 0.07 & 0.39 \\
53226.55 & 8.48 & 0.08 & 0.38 \\
53226.62 & 8.08 & 0.07 & 0.38 \\
53226.83 & 8.42 & 0.08 & 0.35 \\
53226.90 & 8.46 & 0.08 & 0.35 \\
53226.97 & 8.38 & 0.07 & 0.33 \\
53227.04 & 8.08 & 0.07 & 0.35 \\
53227.26 & 8.77 & 0.09 & 0.31 \\
53227.33 & 8.42 & 0.08 & 0.36 \\
53227.40 & 8.31 & 0.07 & 0.34 \\
53227.54 & 8.37 & 0.07 & 0.33 \\
53229.45 & 8.42 & 0.08 & 0.33 \\
53229.94 & 8.38 & 0.07 & 0.36 \\
53230.01 & 8.47 & 0.08 & 0.31 \\
53230.79 & 8.69 & 0.09 & 0.37 \\
53231.21 & 9.49 & 0.12 & 0.35 \\
53231.28 & 9.37 & 0.12 & 0.34 \\
53231.91 & 9.08 & 0.10 & 0.31 \\
    \end{tabular}
\end{minipage}
\hrule
\end{table}


\begin{table}
\scriptsize
\centering
\caption{SMEI light curve data for Nova V2467 Cyg}
\hrule
\begin{minipage}[t]{0.4\linewidth}
	\begin{tabular}[t]{lccc}
Date-Obs MJD & SMEI Mag & SMEI Error & {\bf r}\\
\hline
54160.88 & 13.07 & 0.67 & 0.31 \\
54161.38 & 10.30 & 0.19 & 0.30 \\
54167.16 & 10.63 & 0.22 & 0.42 \\
54170.27 & 9.10 & 0.11 & 0.54 \\
54173.52 & 10.26 & 0.19 & 0.55 \\
54173.80 & 9.14 & 0.11 & 0.57 \\
54173.87 & 8.85 & 0.10 & 0.56 \\
54173.94 & 9.85 & 0.15 & 0.58 \\
54174.01 & 8.78 & 0.09 & 0.55 \\
54174.08 & 7.90 & 0.06 & 0.53 \\
54174.15 & 7.77 & 0.06 & 0.52 \\
54174.22 & 7.77 & 0.06 & 0.58 \\
54174.43 & 7.14 & 0.04 & 0.58 \\
54174.50 & 7.06 & 0.04 & 0.57 \\
54174.57 & 7.12 & 0.04 & 0.63 \\
54174.64 & 6.80 & 0.04 & 0.59 \\
54174.71 & 6.83 & 0.04 & 0.60 \\
54174.79 & 6.80 & 0.04 & 0.62 \\
54174.86 & 6.75 & 0.04 & 0.61 \\
54174.93 & 6.64 & 0.04 & 0.68 \\
54175.00 & 6.65 & 0.04 & 0.65 \\
54175.07 & 6.48 & 0.03 & 0.67 \\
54175.14 & 6.55 & 0.03 & 0.62 \\
54175.21 & 6.61 & 0.04 & 0.63 \\
54175.28 & 6.60 & 0.03 & 0.55 \\
54175.42 & 6.61 & 0.04 & 0.56 \\
54175.49 & 6.50 & 0.03 & 0.67 \\
54175.56 & 6.30 & 0.03 & 0.70 \\
54175.63 & 6.40 & 0.03 & 0.67 \\
    \end{tabular}
\end{minipage}
\begin{minipage}[t]{0.4\linewidth}
	\begin{tabular}[t]{lccc}
Date-Obs MJD & SMEI Mag & SMEI Error  & {\bf r}\\
\hline
54176.48 & 7.21 & 0.05 & 0.35 \\
54176.55 & 7.20 & 0.05 & 0.39 \\
54176.62 & 7.46 & 0.05 & 0.32 \\
54176.69 & 7.20 & 0.05 & 0.37 \\
54176.76 & 7.15 & 0.05 & 0.37 \\
54176.83 & 7.22 & 0.05 & 0.35 \\
54176.90 & 7.27 & 0.05 & 0.35 \\
54181.77 & 9.32 & 0.12 & 0.38 \\
54182.26 & 9.29 & 0.12 & 0.43 \\
54182.48 & 9.18 & 0.11 & 0.37 \\
54182.55 & 9.60 & 0.14 & 0.42 \\
54182.62 & 9.86 & 0.15 & 0.44 \\
54182.69 & 8.55 & 0.08 & 0.43 \\
54182.76 & 9.38 & 0.12 & 0.45 \\
54182.83 & 8.89 & 0.10 & 0.46 \\
54182.90 & 8.44 & 0.08 & 0.43 \\
54183.46 & 10.53 & 0.21 & 0.38 \\
54183.54 & 9.43 & 0.13 & 0.46 \\
54183.61 & 10.44 & 0.20 & 0.40 \\
54183.75 & 8.78 & 0.09 & 0.50 \\
54183.82 & 8.97 & 0.10 & 0.48 \\
54184.03 & 7.73 & 0.06 & 0.33 \\
54184.10 & 7.71 & 0.06 & 0.43 \\
54184.17 & 7.93 & 0.06 & 0.38 \\
54184.59 & 7.66 & 0.06 & 0.31 \\
54184.73 & 8.19 & 0.07 & 0.38 \\
54184.80 & 7.86 & 0.06 & 0.33 \\
54184.88 & 8.17 & 0.07 & 0.37 \\
54184.95 & 7.81 & 0.06 & 0.33 \\
    \end{tabular}
\end{minipage}
\hrule
\end{table}


\begin{table}
\centering
\scriptsize
\caption{SMEI light curve data for Nova V458 Vul}
\hrule
\begin{minipage}[t]{0.4\linewidth}
	\begin{tabular}[t]{lccc}
Date-Obs MJD & SMEI Mag & SMEI Error & {\bf r} \\
\hline
54321.22 & 8.57 & 0.10 & 0.40 \\
54321.86 & 8.06 & 0.08 & 0.62 \\
54321.93 & 8.04 & 0.07 & 0.66 \\
54322.00 & 8.15 & 0.08 & 0.65 \\
54322.07 & 8.16 & 0.08 & 0.65 \\
54322.14 & 8.40 & 0.09 & 0.60 \\
54322.35 & 8.56 & 0.09 & 0.55 \\
54322.49 & 8.36 & 0.09 & 0.53 \\
54322.70 & 8.20 & 0.08 & 0.58 \\
54322.91 & 8.39 & 0.09 & 0.52 \\
54322.98 & 8.38 & 0.09 & 0.53 \\
54323.12 & 8.36 & 0.09 & 0.51 \\
54323.48 & 8.30 & 0.08 & 0.49 \\
54323.55 & 8.33 & 0.09 & 0.49 \\
54323.62 & 8.45 & 0.09 & 0.44 \\
54323.83 & 8.38 & 0.09 & 0.49 \\
54323.90 & 8.45 & 0.09 & 0.47 \\
54324.04 & 8.47 & 0.09 & 0.46 \\
54324.11 & 8.56 & 0.09 & 0.49 \\
54324.18 & 8.20 & 0.08 & 0.56 \\
54324.25 & 8.18 & 0.08 & 0.54 \\
54324.46 & 8.41 & 0.09 & 0.50 \\
54324.54 & 8.50 & 0.09 & 0.47 \\
54324.61 & 8.47 & 0.09 & 0.46 \\
54324.75 & 8.31 & 0.08 & 0.51 \\
54324.82 & 8.51 & 0.09 & 0.47 \\
54324.96 & 8.23 & 0.08 & 0.53 \\
54325.10 & 8.17 & 0.08 & 0.57 \\
54325.17 & 8.14 & 0.08 & 0.58 \\
    \end{tabular}
\end{minipage}
\begin{minipage}[t]{0.4\linewidth}
	\begin{tabular}[t]{lccc}
Date-Obs MJD & SMEI Mag & SMEI Error & {\bf r}  \\
\hline
54325.24 & 8.29 & 0.08 & 0.54 \\
54325.31 & 8.15 & 0.08 & 0.58 \\
54325.38 & 8.11 & 0.08 & 0.56 \\
54325.45 & 8.17 & 0.08 & 0.58 \\
54325.52 & 8.13 & 0.08 & 0.64 \\
54325.59 & 8.07 & 0.08 & 0.64 \\
54325.66 & 7.94 & 0.07 & 0.68 \\
54325.73 & 7.98 & 0.07 & 0.70 \\
54325.80 & 8.00 & 0.07 & 0.67 \\
54325.88 & 7.98 & 0.07 & 0.71 \\
54325.95 & 8.07 & 0.08 & 0.69 \\
54326.02 & 8.13 & 0.08 & 0.67 \\
54326.09 & 8.14 & 0.08 & 0.67 \\
54326.23 & 8.34 & 0.09 & 0.61 \\
54326.30 & 8.43 & 0.09 & 0.55 \\
54326.58 & 8.26 & 0.08 & 0.54 \\
54326.65 & 8.32 & 0.09 & 0.52 \\
54326.72 & 8.47 & 0.09 & 0.48 \\
54326.86 & 8.63 & 0.10 & 0.48 \\
54327.43 & 8.51 & 0.09 & 0.43 \\
54327.57 & 8.57 & 0.10 & 0.48 \\
54328.06 & 8.87 & 0.11 & 0.42 \\
54328.13 & 8.64 & 0.10 & 0.44 \\
54328.20 & 8.31 & 0.08 & 0.47 \\
54328.34 & 8.57 & 0.10 & 0.47 \\
54328.41 & 8.44 & 0.09 & 0.52 \\
54328.63 & 8.84 & 0.11 & 0.48 \\
54328.70 & 8.87 & 0.11 & 0.43 \\
    \end{tabular}
\end{minipage}
\hrule
\end{table}


\begin{table}
\centering
\scriptsize
\caption{SMEI light curve data for Nova V597 Pup}
\hrule
\begin{minipage}[t]{0.4\linewidth}
	\begin{tabular}[t]{lccc}
Date-Obs MJD & SMEI Mag & SMEI Error & {\bf r} \\
\hline
54407.04 & 9.57 & 0.13 & 0.34 \\
54416.43 & 10.48 & 0.20 & 0.43 \\
54416.50 & 10.30 & 0.18 & 0.31 \\
54417.34 & 9.56 & 0.13 & 0.40 \\
54417.48 & 8.79 & 0.09 & 0.37 \\
54417.62 & 8.34 & 0.07 & 0.39 \\
54417.70 & 7.97 & 0.06 & 0.35 \\
\end{tabular}
\end{minipage}
\begin{minipage}[t]{0.4\linewidth}
	\begin{tabular}[t]{lccc}
Date-Obs MJD & SMEI Mag & SMEI Error & {\bf r}\\
\hline
54427.36 & 9.43 & 0.12 & 0.52 \\
54427.43 & 10.06 & 0.16 & 0.48 \\
54427.64 & 10.35 & 0.19 & 0.47 \\
54427.71 & 10.20 & 0.17 & 0.45 \\
54427.86 & 10.73 & 0.22 & 0.44 \\
54428.07 & 11.35 & 0.29 & 0.30 \\
54428.14 & 9.49 & 0.13 & 0.40 \\
\end{tabular}
\end{minipage}
\hrule
\end{table}

\begin{table}
\centering
\scriptsize
\caption{SMEI light curve data for Nova V597 Pup Continued}
\hrule
\begin{minipage}[t]{0.4\linewidth}
	\begin{tabular}[t]{lccc}
Date-Obs MJD & SMEI Mag & SMEI Error & {\bf r} \\
\hline
54447.75 & 9.06 & 0.10 & 0.40 \\
54447.82 & 9.72 & 0.14 & 0.39 \\
54448.04 & 9.43 & 0.12 & 0.35 \\
54448.11 & 9.71 & 0.14 & 0.42 \\
54448.25 & 8.95 & 0.10 & 0.49 \\
54448.46 & 9.43 & 0.12 & 0.39 \\
54449.02 & 9.21 & 0.11 & 0.39 \\
54449.52 & 9.44 & 0.12 & 0.37 \\
54449.59 & 9.79 & 0.14 & 0.41 \\
    \end{tabular}
\end{minipage}
\begin{minipage}[t]{0.4\linewidth}
	\begin{tabular}[t]{lccc}
Date-Obs MJD & SMEI Mag & SMEI Error & {\bf r}\\
\hline
54449.66 & 9.45 & 0.12 & 0.40 \\
54449.87 & 9.38 & 0.12 & 0.40 \\
54450.01 & 9.88 & 0.15 & 0.39 \\
54450.15 & 9.73 & 0.14 & 0.49 \\
54450.36 & 9.48 & 0.13 & 0.42 \\
54450.50 & 10.29 & 0.18 & 0.45 \\
54450.57 & 10.01 & 0.16 & 0.44 \\
54450.64 & 11.04 & 0.26 & 0.42 \\
54450.86 & 10.09 & 0.17 & 0.54 \\
    \end{tabular}
\end{minipage}
\hrule
\end{table}


\begin{table}
\centering
\scriptsize
\caption{SMEI light curve data for Nova V459 Vul}
\hrule
\begin{minipage}[t]{0.4\linewidth}
	\begin{tabular}[t]{lccc}
Date-Obs MJD & SMEI Mag & SMEI Error & {\bf r}\\
\hline
54431.27 & 9.79 & 0.15 & 0.45 \\
54431.91 & 9.02 & 0.11 & 0.41 \\
54449.05 & 8.18 & 0.07 & 0.41 \\
54459.85 & 7.06 & 0.04 & 0.49 \\
54459.99 & 6.86 & 0.04 & 0.48 \\
54461.54 & 6.88 & 0.04 & 0.47 \\
54461.61 & 6.85 & 0.04 & 0.47 \\
54461.68 & 6.79 & 0.04 & 0.52 \\
54461.89 & 6.65 & 0.04 & 0.56 \\
54461.96 & 6.65 & 0.04 & 0.56 \\
54462.11 & 6.59 & 0.04 & 0.58 \\
54462.32 & 6.86 & 0.04 & 0.49 \\
54462.39 & 6.88 & 0.04 & 0.49 \\
54462.46 & 7.52 & 0.05 & 0.44 \\
54462.53 & 6.93 & 0.04 & 0.53 \\
54462.67 & 7.11 & 0.04 & 0.43 \\
54462.74 & 7.00 & 0.04 & 0.43 \\
54462.88 & 6.89 & 0.04 & 0.48 \\
54462.95 & 6.99 & 0.04 & 0.46 \\
54463.02 & 6.88 & 0.04 & 0.53 \\
54463.16 & 6.96 & 0.04 & 0.48 \\
54463.23 & 6.87 & 0.04 & 0.52 \\
54463.45 & 6.84 & 0.04 & 0.50 \\
54463.52 & 6.82 & 0.04 & 0.51 \\
    \end{tabular}
\end{minipage}
\begin{minipage}[t]{0.4\linewidth}
	\begin{tabular}[t]{lccc}
Date-Obs MJD & SMEI Mag & SMEI Error & {\bf r}\\
\hline
54463.73 & 6.82 & 0.04 & 0.57 \\
54463.94 & 6.94 & 0.04 & 0.50 \\
54464.08 & 6.95 & 0.04 & 0.50 \\
54464.29 & 7.05 & 0.04 & 0.50 \\
54464.43 & 7.07 & 0.04 & 0.54 \\
54464.50 & 7.11 & 0.04 & 0.52 \\
54464.57 & 7.19 & 0.05 & 0.51 \\
54464.64 & 7.13 & 0.04 & 0.49 \\
54464.79 & 7.18 & 0.05 & 0.51 \\
54464.93 & 7.46 & 0.05 & 0.47 \\
54465.21 & 7.15 & 0.05 & 0.45 \\
54465.28 & 7.14 & 0.04 & 0.47 \\
54465.35 & 7.14 & 0.04 & 0.50 \\
54466.20 & 7.12 & 0.04 & 0.47 \\
54466.34 & 7.06 & 0.04 & 0.52 \\
54466.41 & 7.15 & 0.05 & 0.48 \\
54466.55 & 7.09 & 0.04 & 0.44 \\
54466.90 & 7.10 & 0.04 & 0.49 \\
54466.97 & 7.14 & 0.05 & 0.54 \\
54467.40 & 7.36 & 0.05 & 0.46 \\
54467.54 & 7.25 & 0.05 & 0.50 \\
54470.79 & 7.27 & 0.05 & 0.48 \\
54477.70 & 8.20 & 0.07 & 0.52 \\
    \end{tabular}
\end{minipage}
\hrule
\end{table}


\begin{table}
\centering
\scriptsize
\caption{SMEI light curve data for Nova V2491 Cyg}
	\begin{tabular}[t]{lccc}
	\hline
Date-Obs MJD & SMEI Mag & SMEI Error & {\bf r} \\
\hline
54566.61 & 7.46 & 0.05 & 0.43 \\
54566.68 & 7.46 & 0.05 & 0.45 \\
54566.75 & 7.42 & 0.05 & 0.45 \\
54566.82 & 7.37 & 0.05 & 0.44 \\
54566.89 & 7.36 & 0.05 & 0.41 \\
54566.96 & 7.54 & 0.05 & 0.46 \\
54567.03 & 7.56 & 0.05 & 0.45 \\
54567.10 & 7.89 & 0.06 & 0.42 \\
54567.17 & 7.90 & 0.06 & 0.40 \\
54567.24 & 7.63 & 0.05 & 0.49 \\
54567.59 & 8.64 & 0.09 & 0.44 \\
54567.66 & 10.67 & 0.22 & 0.31 \\
54567.88 & 12.85 & 0.58 & 0.40 \\
\hline
    \end{tabular}
\end{table}

\begin{table}
\centering
\scriptsize
\caption{SMEI light curve data for Nova QYMus}
\hrule
\begin{minipage}[t]{0.4\linewidth}
	\begin{tabular}[t]{lccc}
Date-Obs MJD & SMEI Mag & SMEI Error  & {\bf r}  \\
\hline
54705.26 & 7.89 & 0.06 & 0.97 \\
54705.40 & 7.89 & 0.06 & 0.97 \\
54705.47 & 7.56 & 0.05 & 0.97 \\
54705.54 & 7.87 & 0.06 & 0.97 \\
54705.61 & 8.33 & 0.08 & 0.97 \\
54705.68 & 7.75 & 0.06 & 0.97 \\
54708.01 & 7.83 & 0.06 & 0.96 \\
\end{tabular}
\end{minipage}
\begin{minipage}[t]{0.4\linewidth}
	\begin{tabular}[t]{lccc}
Date-Obs MJD & SMEI Mag & SMEI Error & {\bf r}\\
\hline
54719.36 & 8.05 & 0.07 & 0.97 \\
54719.43 & 7.86 & 0.06 & 0.97 \\
54719.50 & 8.05 & 0.07 & 0.97 \\
54719.58 & 7.72 & 0.06 & 0.98 \\
54719.65 & 7.94 & 0.06 & 0.97 \\
54720.07 & 8.36 & 0.08 & 0.97 \\
54720.14 & 8.04 & 0.07 & 0.97 \\
\end{tabular}
\end{minipage}
\hrule
\end{table}

\begin{table}
\centering
\scriptsize
\caption{SMEI light curve data for Nova QYMus continued}
\hrule
\begin{minipage}[t]{0.4\linewidth}
	\begin{tabular}[t]{lccc}
Date-Obs MJD & SMEI Mag & SMEI Error & {\bf r} \\
\hline 
54738.34 & 6.95 & 0.04 & 0.97 \\
54738.55 & 7.16 & 0.05 & 0.97 \\
54739.54 & 7.13 & 0.04 & 0.97 \\
54739.68 & 7.29 & 0.05 & 0.97 \\
54743.63 & 8.24 & 0.07 & 0.97 \\
54745.11 & 8.52 & 0.08 & 0.97 \\
54745.39 & 7.96 & 0.07 & 0.95 \\
54745.54 & 8.18 & 0.07 & 0.97 \\
54746.38 & 8.50 & 0.08 & 0.97 \\
54747.23 & 8.65 & 0.09 & 0.97 \\
54748.14 & 8.28 & 0.08 & 0.97 \\
54748.50 & 9.03 & 0.11 & 0.97 \\
54748.64 & 10.54 & 0.21 & 0.97 \\
54748.78 & 8.27 & 0.08 & 0.97 \\
54749.48 & 8.19 & 0.07 & 0.97 \\
54749.70 & 9.04 & 0.11 & 0.96 \\
54750.05 & 8.02 & 0.07 & 0.97 \\
54750.26 & 9.45 & 0.13 & 0.97 \\
54750.68 & 8.20 & 0.07 & 0.96 \\
54751.04 & 8.25 & 0.07 & 0.97 \\
54751.11 & 9.76 & 0.15 & 0.97 \\
54751.25 & 8.20 & 0.07 & 0.97 \\
54751.32 & 7.69 & 0.06 & 0.97 \\
54751.39 & 10.00 & 0.17 & 0.97 \\
54751.74 & 8.24 & 0.07 & 0.97 \\
54751.81 & 7.95 & 0.07 & 0.96 \\
54752.02 & 7.72 & 0.06 & 0.97 \\
54752.30 & 9.00 & 0.11 & 0.97 \\
54752.66 & 8.65 & 0.09 & 0.96 \\
54752.73 & 9.57 & 0.14 & 0.97 \\
54753.01 & 8.51 & 0.08 & 0.97 \\
54753.08 & 9.11 & 0.11 & 0.97 \\
54753.15 & 8.24 & 0.07 & 0.97 \\
54753.22 & 8.03 & 0.07 & 0.97 \\
54753.29 & 8.59 & 0.09 & 0.97 \\
54753.36 & 9.41 & 0.13 & 0.97 \\
54753.58 & 8.68 & 0.09 & 0.97 \\
54754.56 & 8.84 & 0.10 & 0.97 \\
54754.63 & 8.59 & 0.09 & 0.97 \\
54754.85 & 7.48 & 0.05 & 0.92 \\
54755.13 & 9.01 & 0.11 & 0.97 \\
54755.34 & 9.61 & 0.14 & 0.97 \\
54755.48 & 11.97 & 0.41 & 0.97 \\
54755.69 & 8.93 & 0.10 & 0.96 \\
54755.76 & 8.78 & 0.10 & 0.96 \\
54756.40 & 10.11 & 0.17 & 0.96 \\
54756.54 & 9.01 & 0.11 & 0.96 \\
54756.61 & 8.74 & 0.09 & 0.96 \\
54756.68 & 12.22 & 0.46 & 0.96 \\
54756.75 & 8.60 & 0.09 & 0.97 \\
54757.03 & 8.60 & 0.09 & 0.97 \\
54757.17 & 8.35 & 0.08 & 0.97 \\
54757.32 & 9.82 & 0.15 & 0.96 \\
54757.39 & 8.96 & 0.10 & 0.96 \\
54760.84 & 11.27 & 0.30 & 0.95 \\
    \end{tabular}
\end{minipage}
\begin{minipage}[t]{0.4\linewidth}
	\begin{tabular}[t]{lccc}
Date-Obs MJD & SMEI Mag & SMEI Error & {\bf r} \\
\hline 
54761.27 & 8.50 & 0.08 & 0.97 \\
54761.34 & 8.13 & 0.07 & 0.97 \\
54761.41 & 9.00 & 0.11 & 0.97 \\
54761.48 & 8.81 & 0.10 & 0.97 \\
54761.55 & 9.64 & 0.14 & 0.97 \\
54761.62 & 9.11 & 0.11 & 0.97 \\
54761.76 & 9.11 & 0.11 & 0.97 \\
54761.83 & 11.87 & 0.39 & 0.96 \\
54762.04 & 11.77 & 0.37 & 0.97 \\
54762.18 & 9.29 & 0.12 & 0.97 \\
54762.25 & 9.08 & 0.11 & 0.97 \\
54762.32 & 11.02 & 0.26 & 0.97 \\
54762.54 & 8.60 & 0.09 & 0.97 \\
54762.75 & 9.76 & 0.15 & 0.96 \\
54762.82 & 9.31 & 0.12 & 0.97 \\
54763.03 & 9.77 & 0.15 & 0.97 \\
54763.59 & 9.16 & 0.11 & 0.97 \\
54763.66 & 9.68 & 0.14 & 0.97 \\
54763.73 & 9.20 & 0.12 & 0.97 \\
54764.09 & 7.64 & 0.06 & 0.97 \\
54764.16 & 8.36 & 0.08 & 0.97 \\
54764.30 & 8.60 & 0.09 & 0.97 \\
54764.65 & 10.05 & 0.17 & 0.97 \\
54771.70 & 9.59 & 0.14 & 0.96 \\
54772.06 & 8.38 & 0.08 & 0.96 \\
54772.41 & 8.43 & 0.08 & 0.97 \\
54772.48 & 11.49 & 0.33 & 0.97 \\
54772.62 & 9.13 & 0.11 & 0.97 \\
54772.76 & 9.08 & 0.11 & 0.96 \\
54772.83 & 8.36 & 0.08 & 0.96 \\
54772.90 & 10.83 & 0.24 & 0.94 \\
54773.05 & 8.04 & 0.07 & 0.96 \\
54773.12 & 9.00 & 0.11 & 0.96 \\
54773.19 & 8.15 & 0.07 & 0.97 \\
54773.26 & 9.58 & 0.14 & 0.97 \\
54773.33 & 8.68 & 0.09 & 0.96 \\
54773.40 & 8.10 & 0.07 & 0.96 \\
54773.47 & 8.18 & 0.07 & 0.96 \\
54773.54 & 7.86 & 0.06 & 0.96 \\
54773.61 & 8.03 & 0.07 & 0.96 \\
54773.68 & 9.28 & 0.12 & 0.96 \\
54773.75 & 8.22 & 0.07 & 0.96 \\
54773.82 & 10.59 & 0.22 & 0.96 \\
54773.89 & 8.35 & 0.08 & 0.92 \\
54774.03 & 8.25 & 0.07 & 0.96 \\
54774.10 & 8.54 & 0.09 & 0.96 \\
54774.18 & 8.00 & 0.07 & 0.96 \\
54774.25 & 9.53 & 0.13 & 0.96 \\
54774.32 & 7.91 & 0.06 & 0.96 \\
54774.39 & 8.31 & 0.08 & 0.96 \\
54774.46 & 8.35 & 0.08 & 0.97 \\
54774.53 & 9.67 & 0.14 & 0.97 \\
54774.88 & 11.53 & 0.33 & 0.95 \\
54774.95 & 9.23 & 0.12 & 0.92 \\
    \end{tabular}
\end{minipage}
\hrule
\end{table}

\begin{table}
\centering
\scriptsize
\caption{SMEI light curve data for Nova V5583 Sgr}
\hrule
\begin{minipage}[t]{0.4\linewidth}
	\begin{tabular}[t]{lccc}
Date-Obs MJD & SMEI Mag & SMEI Error & {\bf r}\\
\hline
55044.16 & 8.60 & 0.10 & 0.23 \\
55044.23 & 8.29 & 0.08 & 0.19 \\
55044.37 & 8.21 & 0.08 & 0.21 \\
55044.44 & 8.65 & 0.10 & 0.23 \\
55044.58 & 9.88 & 0.17 & 0.20 \\
55048.67 & 8.01 & 0.07 & 0.24 \\
55048.74 & 7.83 & 0.07 & 0.24 \\
55048.81 & 7.77 & 0.07 & 0.21 \\
55048.88 & 7.64 & 0.06 & 0.23 \\
55048.96 & 7.60 & 0.06 & 0.26 \\
55049.10 & 7.69 & 0.06 & 0.26 \\
55049.24 & 7.48 & 0.06 & 0.29 \\
55049.31 & 7.67 & 0.06 & 0.33 \\
55049.38 & 7.32 & 0.05 & 0.31 \\
55049.45 & 7.28 & 0.05 & 0.31 \\
55049.66 & 7.20 & 0.05 & 0.36 \\
55049.73 & 7.18 & 0.05 & 0.37 \\
55049.87 & 7.17 & 0.05 & 0.38 \\
55049.94 & 7.04 & 0.05 & 0.39 \\
55050.01 & 7.08 & 0.05 & 0.38 \\
55050.08 & 6.94 & 0.05 & 0.44 \\
55050.16 & 7.03 & 0.05 & 0.40 \\
55050.23 & 6.98 & 0.05 & 0.42 \\
55050.44 & 7.13 & 0.05 & 0.39 \\
55050.72 & 7.28 & 0.05 & 0.34 \\
55050.79 & 7.34 & 0.05 & 0.32 \\
55051.00 & 7.37 & 0.06 & 0.31 \\
55051.14 & 7.57 & 0.06 & 0.27 \\
55051.21 & 7.55 & 0.06 & 0.28 \\
55051.29 & 7.65 & 0.06 & 0.24 \\
55051.36 & 7.58 & 0.06 & 0.27 \\
55051.43 & 7.57 & 0.06 & 0.28 \\
55051.50 & 7.54 & 0.06 & 0.27 \\
55051.57 & 7.62 & 0.06 & 0.26 \\
55051.64 & 7.58 & 0.06 & 0.26 \\
    \end{tabular}
\end{minipage}
\begin{minipage}[t]{0.4\linewidth}
	\begin{tabular}[t]{lccc}
Date-Obs MJD & SMEI Mag & SMEI Error & {\bf r}\\
\hline
55051.71 & 7.61 & 0.06 & 0.26 \\
55051.78 & 7.53 & 0.06 & 0.24 \\
55051.85 & 7.58 & 0.06 & 0.23 \\
55051.92 & 7.63 & 0.06 & 0.25 \\
55051.99 & 7.55 & 0.06 & 0.25 \\
55052.06 & 7.56 & 0.06 & 0.24 \\
55052.13 & 7.62 & 0.06 & 0.24 \\
55052.20 & 7.62 & 0.06 & 0.23 \\
55052.27 & 7.44 & 0.06 & 0.26 \\
55052.34 & 7.53 & 0.06 & 0.25 \\
55052.41 & 7.56 & 0.06 & 0.25 \\
55052.48 & 7.56 & 0.06 & 0.27 \\
55052.55 & 7.48 & 0.06 & 0.28 \\
55052.70 & 7.50 & 0.06 & 0.29 \\
55052.77 & 7.50 & 0.06 & 0.30 \\
55053.26 & 7.91 & 0.07 & 0.29 \\
55053.33 & 7.84 & 0.07 & 0.28 \\
55053.40 & 7.93 & 0.07 & 0.27 \\
55053.47 & 7.98 & 0.07 & 0.25 \\
55053.54 & 7.91 & 0.07 & 0.25 \\
55053.61 & 7.91 & 0.07 & 0.24 \\
55053.68 & 8.00 & 0.07 & 0.21 \\
55053.75 & 8.11 & 0.08 & 0.20 \\
55053.82 & 7.90 & 0.07 & 0.19 \\
55053.89 & 8.12 & 0.08 & 0.17 \\
55053.96 & 8.29 & 0.08 & 0.14 \\
55054.04 & 8.09 & 0.08 & 0.16 \\
55054.11 & 7.92 & 0.07 & 0.18 \\
55054.18 & 8.34 & 0.09 & 0.15 \\
55054.25 & 8.19 & 0.08 & 0.16 \\
55054.32 & 8.05 & 0.08 & 0.19 \\
55054.39 & 7.88 & 0.07 & 0.22 \\
55054.53 & 7.85 & 0.07 & 0.24 \\
55054.60 & 8.34 & 0.09 & 0.24 \\
55054.67 & 8.21 & 0.08 & 0.26 \\
    \end{tabular}
\end{minipage}
\hrule
\end{table}

\begin{table}
\scriptsize
\caption{SMEI light curve data for Nova RS Oph}
\hrule
\begin{minipage}[t]{0.333\linewidth}
	\begin{tabular}[t]{lcc}
Date-Obs MJD & SMEI Mag & SMEI Error \\
\hline
53775.77	&	7.79	&	0.06	\\
53775.84	&	7.42	&	0.05	\\
53776.26	&	7.07	&	0.04	\\
53778.03	&	6.18	&	0.03	\\
53778.10	&	5.66	&	0.02	\\
53778.17	&	5.30	&	0.02	\\
53778.24	&	4.93	&	0.02	\\
53778.31	&	4.59	&	0.01	\\
53778.38	&	4.44	&	0.01	\\
53778.45	&	4.46	&	0.01	\\
53778.52	&	4.36	&	0.01	\\
53778.59	&	4.20	&	0.01	\\
53778.66	&	4.09	&	0.01	\\
53778.73	&	3.97	&	0.01	\\
53778.80	&	3.93	&	0.01	\\
53778.87	&	3.93	&	0.01	\\
53778.94	&	3.87	&	0.01	\\
53779.01	&	3.93	&	0.01	\\
53779.09	&	3.98	&	0.01	\\
53779.16	&	4.03	&	0.01	\\
53779.23	&	4.12	&	0.01	\\
53779.30	&	4.16	&	0.01	\\
53779.37	&	4.23	&	0.01	\\
53779.44	&	4.26	&	0.01	\\
53779.51	&	4.29	&	0.01	\\
53779.58	&	4.39	&	0.01	\\
53779.65	&	4.38	&	0.01	\\
53779.72	&	4.46	&	0.01	\\
53779.79	&	4.47	&	0.01	\\
53779.86	&	4.51	&	0.01	\\
53779.93	&	4.56	&	0.01	\\
53780.00	&	4.59	&	0.01	\\
53780.07	&	4.59	&	0.01	\\
53780.14	&	4.59	&	0.01	\\
53780.21	&	4.63	&	0.01	\\
53780.28	&	4.63	&	0.01	\\
53780.36	&	4.62	&	0.01	\\
53780.43	&	4.67	&	0.01	\\
53780.50	&	4.67	&	0.01	\\
53780.57	&	4.66	&	0.01	\\
53780.64	&	4.73	&	0.01	\\
53780.71	&	4.72	&	0.01	\\
53780.78	&	4.73	&	0.01	\\
53780.85	&	4.71	&	0.01	\\
53780.92	&	4.73	&	0.01	\\
53781.27	&	4.91	&	0.02	\\
53781.34	&	4.88	&	0.02	\\
    \end{tabular}
\end{minipage}
\begin{minipage}[t]{0.333\linewidth}
	\begin{tabular}[t]{lcc}
Date-Obs MJD & SMEI Mag & SMEI Error \\
\hline
53781.41	&	4.85	&	0.02	\\
53781.48	&	4.90	&	0.02	\\
53781.55	&	4.96	&	0.02	\\
53781.63	&	4.99	&	0.02	\\
53781.70	&	4.95	&	0.02	\\
53781.77	&	5.00	&	0.02	\\
53781.84	&	5.00	&	0.02	\\
53781.91	&	4.93	&	0.02	\\
53781.98	&	5.02	&	0.02	\\
53782.05	&	5.00	&	0.02	\\
53782.12	&	5.11	&	0.02	\\
53782.19	&	5.06	&	0.02	\\
53782.26	&	5.13	&	0.02	\\
53782.75	&	5.19	&	0.02	\\
53782.82	&	5.26	&	0.02	\\
53782.90	&	5.25	&	0.02	\\
53782.97	&	5.30	&	0.02	\\
53783.04	&	5.29	&	0.02	\\
53783.11	&	5.30	&	0.02	\\
53783.18	&	5.33	&	0.02	\\
53783.25	&	5.32	&	0.02	\\
53783.32	&	5.36	&	0.02	\\
53783.39	&	5.27	&	0.02	\\
53783.53	&	5.34	&	0.02	\\
53783.60	&	5.33	&	0.02	\\
53783.67	&	5.35	&	0.02	\\
53783.74	&	5.38	&	0.02	\\
53783.81	&	5.36	&	0.02	\\
53783.88	&	5.38	&	0.02	\\
53783.95	&	5.38	&	0.02	\\
53784.02	&	5.37	&	0.02	\\
53784.10	&	5.47	&	0.02	\\
53784.17	&	5.47	&	0.02	\\
53784.24	&	5.38	&	0.02	\\
53784.31	&	5.56	&	0.02	\\
53784.38	&	5.45	&	0.02	\\
53784.45	&	5.46	&	0.02	\\
53784.52	&	5.48	&	0.02	\\
53784.59	&	5.51	&	0.02	\\
53784.66	&	5.53	&	0.02	\\
53784.73	&	5.53	&	0.02	\\
53784.80	&	5.59	&	0.02	\\
53784.87	&	5.60	&	0.02	\\
53784.94	&	5.67	&	0.02	\\
53785.01	&	5.60	&	0.02	\\
53785.08	&	5.71	&	0.02	\\
53785.15	&	5.65	&	0.02	\\
    \end{tabular}
\end{minipage}
\begin{minipage}[t]{0.333\linewidth}
	\begin{tabular}[t]{lcc}
Date-Obs MJD & SMEI Mag & SMEI Error \\
\hline
53785.22	&	5.72	&	0.02	\\
53785.29	&	5.72	&	0.02	\\
53785.37	&	5.66	&	0.02	\\
53785.44	&	5.66	&	0.02	\\
53785.51	&	5.67	&	0.02	\\
53785.58	&	5.72	&	0.02	\\
53785.65	&	5.72	&	0.02	\\
53785.72	&	5.63	&	0.02	\\
53785.79	&	5.64	&	0.02	\\
53785.86	&	5.59	&	0.02	\\
53785.93	&	5.63	&	0.02	\\
53786.00	&	5.73	&	0.02	\\
53786.07	&	5.75	&	0.02	\\
53786.14	&	5.75	&	0.02	\\
53786.21	&	5.77	&	0.02	\\
53786.28	&	5.77	&	0.02	\\
53786.35	&	5.75	&	0.02	\\
53786.42	&	5.74	&	0.02	\\
53786.49	&	5.79	&	0.02	\\
53786.56	&	5.78	&	0.02	\\
53786.78	&	5.77	&	0.02	\\
53786.85	&	5.83	&	0.02	\\
53786.92	&	5.80	&	0.02	\\
53786.99	&	5.84	&	0.02	\\
53787.06	&	5.80	&	0.02	\\
53787.20	&	5.82	&	0.02	\\
53787.27	&	5.88	&	0.03	\\
53787.34	&	5.93	&	0.03	\\
53787.41	&	5.91	&	0.03	\\
53787.48	&	5.98	&	0.03	\\
53787.55	&	5.89	&	0.03	\\
53787.69	&	6.00	&	0.03	\\
53787.76	&	6.09	&	0.03	\\
53787.83	&	5.95	&	0.03	\\
53787.91	&	6.08	&	0.03	\\
53787.98	&	6.00	&	0.03	\\
53790.16	&	6.24	&	0.03	\\
53790.37	&	6.11	&	0.03	\\
53790.45	&	6.21	&	0.03	\\
53794.47	&	6.23	&	0.03	\\
53794.68	&	6.05	&	0.03	\\
53794.75	&	6.10	&	0.03	\\
53794.82	&	6.46	&	0.03	\\
53794.89	&	6.48	&	0.03	\\
53794.96	&	6.15	&	0.03	\\
    \end{tabular}
\end{minipage}
\hrule
\end{table}


\begin{table}
\scriptsize
\caption{SMEI light curve data for Nova V1280 Sco}
\hrule
\begin{minipage}[t]{0.333\linewidth}
	\begin{tabular}[t]{lcc}
Date-Obs MJD & SMEI Mag & SMEI Error \\
\hline
54144.61	&	6.19	&	0.03	\\
54144.68	&	6.18	&	0.03	\\
54144.75	&	6.20	&	0.03	\\
54144.82	&	6.19	&	0.03	\\
54144.89	&	6.16	&	0.03	\\
54144.96	&	6.16	&	0.03	\\
54145.24	&	6.14	&	0.03	\\
54145.31	&	6.20	&	0.03	\\
54145.38	&	6.10	&	0.03	\\
54145.52	&	6.02	&	0.03	\\
54145.59	&	5.89	&	0.02	\\
54145.66	&	5.81	&	0.02	\\
54145.73	&	5.70	&	0.02	\\
54145.80	&	5.59	&	0.02	\\
54145.88	&	5.48	&	0.02	\\
54145.95	&	5.37	&	0.02	\\
54146.02	&	5.30	&	0.02	\\
54146.09	&	5.27	&	0.02	\\
54146.23	&	5.23	&	0.02	\\
54146.30	&	5.24	&	0.02	\\
54146.37	&	5.23	&	0.02	\\
54146.44	&	5.22	&	0.02	\\
54146.51	&	5.24	&	0.02	\\
54146.58	&	5.22	&	0.02	\\
54146.65	&	5.23	&	0.02	\\
54146.72	&	5.14	&	0.02	\\
54146.79	&	4.75	&	0.01	\\
54146.86	&	4.37	&	0.01	\\
54146.93	&	4.12	&	0.01	\\
54147.00	&	4.05	&	0.01	\\
54147.07	&	4.00	&	0.01	\\
54147.15	&	4.00	&	0.01	\\
54147.22	&	4.02	&	0.01	\\
54147.29	&	4.04	&	0.01	\\
54147.36	&	4.14	&	0.01	\\
54147.43	&	4.13	&	0.01	\\
54147.50	&	4.19	&	0.01	\\
54147.57	&	4.25	&	0.01	\\
54147.64	&	4.41	&	0.01	\\
54147.71	&	4.54	&	0.01	\\
54147.78	&	4.68	&	0.01	\\
54147.85	&	4.79	&	0.01	\\
54147.92	&	4.85	&	0.02	\\
54148.06	&	4.85	&	0.02	\\
54148.13	&	4.73	&	0.01	\\
54148.27	&	4.35	&	0.01	\\
54148.34	&	4.23	&	0.01	\\
54148.41	&	4.24	&	0.01	\\
54148.49	&	4.25	&	0.01	\\
54148.56	&	4.29	&	0.01	\\
54148.63	&	4.33	&	0.01	\\
54148.70	&	4.41	&	0.01	\\
54148.77	&	4.49	&	0.01	\\
54148.84	&	4.56	&	0.01	\\
54148.91	&	4.64	&	0.01	\\
54148.98	&	4.68	&	0.01	\\
54149.05	&	4.71	&	0.01	\\
54149.12	&	4.71	&	0.01	\\
54149.19	&	4.74	&	0.01	\\
54149.26	&	4.76	&	0.01	\\
54149.33	&	4.79	&	0.01	\\
54149.40	&	4.81	&	0.01	\\
54149.47	&	4.80	&	0.01	\\
54149.54	&	4.77	&	0.01	\\
54149.61	&	4.74	&	0.01	\\
54149.68	&	4.70	&	0.01	\\
54149.76	&	4.64	&	0.01	\\
    \end{tabular}
\end{minipage}
\begin{minipage}[t]{0.333\linewidth}
	\begin{tabular}[t]{lcc}
Date-Obs MJD & SMEI Mag & SMEI Error \\
\hline
54149.83	&	4.56	&	0.01	\\
54149.90	&	4.46	&	0.01	\\
54149.97	&	4.34	&	0.01	\\
54150.04	&	4.25	&	0.01	\\
54150.11	&	4.16	&	0.01	\\
54150.18	&	4.13	&	0.01	\\
54150.25	&	4.13	&	0.01	\\
54150.32	&	4.15	&	0.01	\\
54150.39	&	4.22	&	0.01	\\
54150.53	&	4.44	&	0.01	\\
54150.60	&	4.57	&	0.01	\\
54150.67	&	4.74	&	0.01	\\
54150.74	&	4.92	&	0.02	\\
54150.81	&	4.94	&	0.02	\\
54150.88	&	4.89	&	0.02	\\
54150.95	&	4.85	&	0.02	\\
54151.03	&	4.81	&	0.01	\\
54151.10	&	4.81	&	0.01	\\
54151.17	&	4.86	&	0.02	\\
54151.24	&	4.85	&	0.02	\\
54151.31	&	4.86	&	0.02	\\
54151.38	&	4.80	&	0.01	\\
54151.45	&	4.73	&	0.01	\\
54151.52	&	4.68	&	0.01	\\
54151.59	&	4.67	&	0.01	\\
54151.66	&	4.68	&	0.01	\\
54151.73	&	4.71	&	0.01	\\
54151.80	&	4.71	&	0.01	\\
54151.87	&	4.71	&	0.01	\\
54151.94	&	4.69	&	0.01	\\
54152.22	&	4.78	&	0.01	\\
54152.30	&	4.75	&	0.01	\\
54152.37	&	4.77	&	0.01	\\
54152.44	&	4.77	&	0.01	\\
54152.51	&	4.73	&	0.01	\\
54152.58	&	4.65	&	0.01	\\
54152.65	&	4.70	&	0.01	\\
54152.72	&	4.75	&	0.01	\\
54152.79	&	4.80	&	0.01	\\
54152.86	&	4.87	&	0.02	\\
54152.93	&	4.89	&	0.02	\\
54153.00	&	4.92	&	0.02	\\
54153.07	&	4.90	&	0.02	\\
54153.14	&	4.85	&	0.02	\\
54153.21	&	4.77	&	0.01	\\
54153.28	&	4.77	&	0.01	\\
54153.35	&	4.77	&	0.01	\\
54153.42	&	4.80	&	0.01	\\
54153.49	&	4.83	&	0.02	\\
54153.56	&	4.84	&	0.02	\\
54153.64	&	4.86	&	0.02	\\
54153.71	&	4.85	&	0.02	\\
54153.78	&	4.86	&	0.02	\\
54153.85	&	4.88	&	0.02	\\
54153.92	&	4.87	&	0.02	\\
54154.06	&	4.84	&	0.02	\\
54154.13	&	4.84	&	0.02	\\
54154.20	&	4.83	&	0.02	\\
54154.27	&	4.83	&	0.02	\\
54154.34	&	4.87	&	0.02	\\
54154.41	&	4.88	&	0.02	\\
54154.48	&	4.86	&	0.02	\\
54154.55	&	4.87	&	0.02	\\
54154.62	&	4.86	&	0.02	\\
54154.69	&	4.87	&	0.02	\\
54154.76	&	4.89	&	0.02	\\
54154.83	&	4.93	&	0.02	\\
    \end{tabular}
\end{minipage}
\begin{minipage}[t]{0.333\linewidth}
	\begin{tabular}[t]{lcc}
Date-Obs MJD & SMEI Mag & SMEI Error \\
\hline
54154.91	&	4.98	&	0.02	\\
54154.98	&	4.97	&	0.02	\\
54155.05	&	4.98	&	0.02	\\
54155.12	&	5.00	&	0.02	\\
54155.19	&	4.98	&	0.02	\\
54155.26	&	4.96	&	0.02	\\
54155.33	&	4.97	&	0.02	\\
54155.40	&	4.98	&	0.02	\\
54155.47	&	5.02	&	0.02	\\
54155.54	&	5.04	&	0.02	\\
54155.61	&	5.06	&	0.02	\\
54155.68	&	5.08	&	0.02	\\
54155.75	&	5.10	&	0.02	\\
54155.82	&	5.09	&	0.02	\\
54155.89	&	5.05	&	0.02	\\
54155.96	&	5.07	&	0.02	\\
54156.03	&	5.03	&	0.02	\\
54156.10	&	5.03	&	0.02	\\
54156.18	&	5.04	&	0.02	\\
54156.25	&	5.03	&	0.02	\\
54156.32	&	5.03	&	0.02	\\
54156.74	&	5.04	&	0.02	\\
54156.81	&	5.03	&	0.02	\\
54157.16	&	5.04	&	0.02	\\
54157.23	&	5.06	&	0.02	\\
54157.30	&	5.11	&	0.02	\\
54157.37	&	5.14	&	0.02	\\
54157.45	&	5.15	&	0.02	\\
54157.52	&	5.15	&	0.02	\\
54157.59	&	5.16	&	0.02	\\
54157.66	&	5.20	&	0.02	\\
54157.73	&	5.22	&	0.02	\\
54157.80	&	5.25	&	0.02	\\
54157.87	&	5.25	&	0.02	\\
54157.94	&	5.29	&	0.02	\\
54158.01	&	5.30	&	0.02	\\
54158.08	&	5.32	&	0.02	\\
54158.15	&	5.35	&	0.02	\\
54158.22	&	5.40	&	0.02	\\
54158.29	&	5.42	&	0.02	\\
54158.36	&	5.44	&	0.02	\\
54158.43	&	5.48	&	0.02	\\
54158.50	&	5.49	&	0.02	\\
54158.57	&	5.53	&	0.02	\\
54158.64	&	5.51	&	0.02	\\
54158.72	&	5.53	&	0.02	\\
54158.79	&	5.57	&	0.02	\\
54158.86	&	5.55	&	0.02	\\
54158.93	&	5.60	&	0.02	\\
54159.28	&	5.63	&	0.02	\\
54159.35	&	5.73	&	0.02	\\
54159.42	&	5.81	&	0.02	\\
54159.49	&	5.87	&	0.02	\\
54159.56	&	5.92	&	0.02	\\
54159.63	&	5.95	&	0.03	\\
54159.70	&	5.94	&	0.03	\\
54159.77	&	5.95	&	0.03	\\
54159.84	&	5.99	&	0.03	\\
54159.98	&	6.00	&	0.03	\\
54160.13	&	6.07	&	0.03	\\
54160.27	&	6.19	&	0.03	\\
54160.41	&	6.17	&	0.03	\\
54160.48	&	6.18	&	0.03	\\
54160.55	&	6.22	&	0.03	\\
54160.62	&	6.23	&	0.03	\\
54160.69	&	6.24	&	0.03	\\
54160.76	&	6.26	&	0.03	\\
    \end{tabular}
\end{minipage}
\hrule
\end{table}

\begin{table}
\scriptsize
\caption{SMEI light curve data for Nova V1280 Sco continued}
\hrule
\begin{minipage}[t]{0.333\linewidth}
	\begin{tabular}[t]{lcc}
Date-Obs MJD & SMEI Mag & SMEI Error \\
\hline
54160.83	&	6.25	&	0.03	\\
54160.90	&	6.27	&	0.03	\\
54160.97	&	6.34	&	0.03	\\
54161.04	&	6.39	&	0.03	\\
54161.11	&	6.42	&	0.03	\\
54161.18	&	6.35	&	0.03	\\
54161.40	&	6.45	&	0.03	\\
54161.47	&	6.49	&	0.03	\\
54161.54	&	6.53	&	0.03	\\
54161.61	&	6.63	&	0.03	\\
54161.68	&	6.66	&	0.03	\\
54161.75	&	6.61	&	0.03	\\
54161.82	&	6.69	&	0.04	\\
54161.89	&	6.79	&	0.04	\\
54161.96	&	6.95	&	0.04	\\
54162.03	&	6.96	&	0.04	\\
54162.10	&	7.03	&	0.04	\\
    \end{tabular}
\end{minipage}
\begin{minipage}[t]{0.333\linewidth}
	\begin{tabular}[t]{lcc}
Date-Obs MJD & SMEI Mag & SMEI Error \\
\hline
54162.17	&	7.10	&	0.04	\\
54162.24	&	7.12	&	0.04	\\
54162.31	&	7.20	&	0.04	\\
54162.38	&	7.14	&	0.04	\\
54162.45	&	7.20	&	0.04	\\
54162.52	&	7.30	&	0.05	\\
54162.60	&	7.32	&	0.05	\\
54162.67	&	7.42	&	0.05	\\
54162.74	&	7.55	&	0.05	\\
54162.81	&	7.55	&	0.05	\\
54162.88	&	7.55	&	0.05	\\
54162.95	&	7.57	&	0.05	\\
54163.02	&	7.80	&	0.06	\\
54163.09	&	7.60	&	0.05	\\
54163.16	&	7.69	&	0.06	\\
54163.23	&	7.79	&	0.06	\\
54163.30	&	7.61	&	0.05	\\
    \end{tabular}
\end{minipage}
\begin{minipage}[t]{0.333\linewidth}
	\begin{tabular}[t]{lcc}
Date-Obs MJD & SMEI Mag & SMEI Error \\
\hline
54163.37	&	8.21	&	0.07	\\
54163.44	&	8.19	&	0.07	\\
54163.51	&	8.04	&	0.06	\\
54163.58	&	8.29	&	0.07	\\
54163.65	&	8.10	&	0.07	\\
54163.72	&	8.22	&	0.07	\\
54163.79	&	8.31	&	0.07	\\
54163.87	&	8.69	&	0.09	\\
54163.94	&	9.08	&	0.10	\\
54164.01	&	8.86	&	0.09	\\
54164.08	&	9.78	&	0.14	\\
54164.15	&	9.41	&	0.12	\\
54164.22	&	10.02	&	0.16	\\
54164.29	&	9.72	&	0.14	\\
54164.36	&	11.18	&	0.27	\\
54164.43	&	10.21	&	0.17	\\
54164.71	&	10.70	&	0.22	\\
    \end{tabular}
\end{minipage}
\hrule
\end{table}


\begin{table}
\scriptsize
\caption{SMEI light curve data for Nova V598 Pup}
\hrule
\begin{minipage}[t]{0.333\linewidth}
	\begin{tabular}[t]{lcc}
Date-Obs MJD & SMEI Mag & SMEI Error \\
\hline
54248.82	&	9.60	&	0.14	\\	
54254.47	&	7.54	&	0.05	\\	
54254.61	&	6.36	&	0.03	\\	
54254.68	&	6.06	&	0.03	\\	
54254.75	&	5.88	&	0.03	\\	
54254.82	&	5.57	&	0.02	\\	
54254.96	&	4.94	&	0.02	\\	
54255.03	&	5.08	&	0.02	\\	
54255.10	&	5.24	&	0.02	\\	
54255.17	&	5.13	&	0.02	\\	
54255.24	&	5.05	&	0.02	\\	
54255.31	&	4.77	&	0.02	\\	
54255.38	&	4.72	&	0.01	\\	
54255.45	&	4.62	&	0.01	\\	
54255.52	&	4.49	&	0.01	\\	
54255.59	&	4.34	&	0.01	\\	
54255.67	&	4.38	&	0.01	\\	
54255.74	&	4.29	&	0.01	\\	
54255.81	&	4.04	&	0.01	\\	
54255.88	&	4.06	&	0.01	\\	
54255.95	&	3.90	&	0.01	\\	
54256.02	&	3.97	&	0.01	\\	
54256.09	&	3.92	&	0.01	\\	
54256.16	&	3.87	&	0.01	\\	
54256.23	&	3.68	&	0.01	\\	
54256.30	&	3.62	&	0.01	\\	
54256.65	&	3.49	&	0.01	\\	
54256.72	&	3.50	&	0.01	\\	
54256.79	&	3.67	&	0.01	\\	
54256.94	&	3.48	&	0.01	\\	
54257.29	&	3.46	&	0.01	\\
54257.36	&	3.57	&	0.01	\\
54257.43	&	3.48	&	0.01	\\
54257.50	&	3.52	&	0.01	\\
54257.57	&	3.53	&	0.01	\\
54257.64	&	3.57	&	0.01	\\
54257.71	&	3.62	&	0.01	\\
54257.78	&	3.61	&	0.01	\\
54257.99	&	3.81	&	0.01	\\
54258.06	&	3.82	&	0.01	\\
    \end{tabular}
\end{minipage}
\begin{minipage}[t]{0.333\linewidth}
	\begin{tabular}[t]{lcc}
Date-Obs MJD & SMEI Mag & SMEI Error \\
\hline
54258.13	&	3.84	&	0.01	\\
54258.20	&	3.84	&	0.01	\\
54258.28	&	3.89	&	0.01	\\
54258.35	&	3.95	&	0.01	\\
54258.42	&	3.99	&	0.01	\\
54258.49	&	4.04	&	0.01	\\
54258.56	&	4.19	&	0.01	\\
54258.63	&	4.08	&	0.01	\\
54258.70	&	4.10	&	0.01	\\
54258.77	&	4.15	&	0.01	\\
54258.98	&	4.27	&	0.01	\\
54259.05	&	4.28	&	0.01	\\
54259.12	&	4.33	&	0.01	\\
54259.19	&	4.34	&	0.01	\\
54259.26	&	4.41	&	0.01	\\
54259.33	&	4.43	&	0.01	\\
54279.71	&	7.04	&	0.04	\\
54279.79	&	7.07	&	0.04	\\
54279.86	&	6.87	&	0.04	\\
54280.07	&	7.10	&	0.04	\\
54280.14	&	7.19	&	0.05	\\
54280.21	&	7.13	&	0.04	\\
54280.28	&	6.98	&	0.04	\\
54280.35	&	7.19	&	0.05	\\
54280.42	&	7.18	&	0.05	\\
54280.49	&	7.32	&	0.05	\\
54280.56	&	7.19	&	0.05	\\
54280.63	&	7.27	&	0.05	\\
54280.70	&	7.31	&	0.05	\\
54280.77	&	7.39	&	0.05	\\
54280.84	&	7.14	&	0.04	\\
54281.34	&	7.23	&	0.05	\\
54281.41	&	7.81	&	0.06	\\
54281.48	&	7.91	&	0.06	\\
54281.55	&	8.20	&	0.07	\\
54281.62	&	7.84	&	0.06	\\
54281.69	&	7.54	&	0.05	\\
54281.76	&	7.88	&	0.06	\\
54282.04	&	9.53	&	0.13	\\
54282.18	&	8.63	&	0.09	\\
    \end{tabular}
\end{minipage}
\begin{minipage}[t]{0.333\linewidth}
	\begin{tabular}[t]{lcc}
Date-Obs MJD & SMEI Mag & SMEI Error \\
\hline
54282.32	&	10.65	&	0.22	\\
54282.54	&	9.38	&	0.12	\\
54282.68	&	9.57	&	0.14	\\
54282.82	&	9.28	&	0.12	\\
54282.89	&	11.51	&	0.33	\\
54283.38	&	7.16	&	0.05	\\
54283.45	&	7.19	&	0.05	\\
54283.52	&	8.90	&	0.10	\\
54283.59	&	10.36	&	0.19	\\
54283.66	&	7.03	&	0.04	\\
54283.73	&	7.18	&	0.05	\\
54283.80	&	6.85	&	0.04	\\
54286.84	&	7.40	&	0.05	\\
54286.91	&	7.55	&	0.05	\\
54287.05	&	7.75	&	0.06	\\
54287.12	&	8.43	&	0.08	\\
54287.19	&	7.96	&	0.07	\\
54287.26	&	8.24	&	0.07	\\
54287.33	&	8.02	&	0.07	\\
54287.40	&	8.23	&	0.07	\\
54287.47	&	8.27	&	0.08	\\
54287.54	&	8.02	&	0.07	\\
54287.61	&	9.09	&	0.11	\\
54287.68	&	8.16	&	0.07	\\
54287.75	&	8.21	&	0.07	\\
54288.18	&	7.97	&	0.07	\\
54288.25	&	7.95	&	0.06	\\
54288.32	&	7.64	&	0.06	\\
54288.46	&	7.50	&	0.05	\\
54288.53	&	8.01	&	0.07	\\
54288.60	&	8.94	&	0.10	\\
54288.67	&	8.71	&	0.09	\\
54288.74	&	7.76	&	0.06	\\
54288.81	&	7.72	&	0.06	\\
54288.88	&	7.64	&	0.06	\\
54288.95	&	7.92	&	0.06	\\
54289.16	&	8.46	&	0.08	\\
54289.23	&	8.36	&	0.08	\\
54289.31	&	8.52	&	0.08	\\
54289.38	&	7.90	&	0.06	\\
    \end{tabular}
\end{minipage}
\hrule
\end{table}

\begin{table}
\scriptsize
\caption{SMEI light curve data for Nova V598 Pup continued}
\hrule
\begin{minipage}[t]{0.333\linewidth}
	\begin{tabular}[t]{lcc}
Date-Obs MJD & SMEI Mag & SMEI Error \\
\hline
54289.45	&	8.44	&	0.08	\\
54289.52	&	8.73	&	0.09	\\
54289.66	&	8.12	&	0.07	\\
54289.73	&	7.96	&	0.07	\\
54289.80	&	7.81	&	0.06	\\
54289.87	&	8.31	&	0.08	\\
54289.94	&	7.99	&	0.07	\\
54290.01	&	7.67	&	0.06	\\
54290.08	&	7.73	&	0.06	\\
54290.15	&	7.76	&	0.06	\\
54292.90	&	8.37	&	0.08	\\
54292.97	&	7.67	&	0.06	\\
54293.04	&	7.73	&	0.06	\\
54293.11	&	7.72	&	0.06	\\
54293.18	&	7.69	&	0.06	\\
54293.25	&	7.83	&	0.06	\\
54293.33	&	7.32	&	0.05	\\
54293.40	&	7.81	&	0.06	\\
54293.47	&	7.77	&	0.06	\\
54293.54	&	7.78	&	0.06	\\
54293.61	&	7.48	&	0.05	\\
54293.68	&	7.53	&	0.05	\\
54293.75	&	7.93	&	0.06	\\
54293.82	&	7.89	&	0.06	\\
54293.89	&	7.55	&	0.05	\\
54293.96	&	8.06	&	0.07	\\
54297.06	&	12.19	&	0.45	\\
54297.84	&	7.82	&	0.06	\\
54297.91	&	7.41	&	0.05	\\
54297.98	&	8.18	&	0.07	\\
54298.05	&	8.76	&	0.09	\\
54298.12	&	7.53	&	0.05	\\
54298.19	&	7.23	&	0.05	\\
54298.26	&	12.24	&	0.46	\\
54298.33	&	7.75	&	0.06	\\
54298.40	&	7.92	&	0.06	\\
54298.47	&	7.97	&	0.07	\\
54298.54	&	7.92	&	0.06	\\
54298.62	&	6.70	&	0.04	\\
54298.76	&	8.19	&	0.07	\\
54298.83	&	7.06	&	0.04	\\
54303.20	&	13.64	&	0.86	\\
54306.73	&	7.77	&	0.06	\\
54306.80	&	9.09	&	0.11	\\
54306.87	&	7.95	&	0.06	\\
54306.94	&	8.17	&	0.07	\\
54307.01	&	8.71	&	0.09	\\
54307.15	&	9.80	&	0.15	\\
54307.22	&	9.41	&	0.13	\\
54307.29	&	9.06	&	0.11	\\
54307.36	&	8.29	&	0.08	\\
54307.43	&	9.76	&	0.15	\\
54307.50	&	8.76	&	0.09	\\
54307.57	&	9.11	&	0.11	\\
54307.64	&	9.55	&	0.13	\\
54307.71	&	10.08	&	0.17	\\
54307.78	&	9.05	&	0.11	\\
54308.00	&	10.45	&	0.20	\\
54308.07	&	8.37	&	0.08	\\
54308.14	&	7.68	&	0.06	\\
54308.28	&	9.97	&	0.16	\\
54308.35	&	7.61	&	0.06	\\
54308.42	&	7.39	&	0.05	\\
54308.49	&	10.21	&	0.18	\\
54308.56	&	7.47	&	0.05	\\
54308.63	&	9.67	&	0.14	\\
54308.70	&	6.45	&	0.03	\\
54308.84	&	8.93	&	0.10	\\
54308.92	&	7.53	&	0.05	\\
54308.98	&	7.94	&	0.06	\\
    \end{tabular}
\end{minipage}
\begin{minipage}[t]{0.333\linewidth}
	\begin{tabular}[t]{lcc}
Date-Obs MJD & SMEI Mag & SMEI Error \\
\hline
54309.27	&	8.23	&	0.07	\\
54309.34	&	8.28	&	0.08	\\
54309.41	&	8.00	&	0.07	\\
54309.48	&	7.79	&	0.06	\\
54309.55	&	7.87	&	0.06	\\
54309.62	&	7.84	&	0.06	\\
54309.69	&	10.94	&	0.25	\\
54309.76	&	7.68	&	0.06	\\
54309.83	&	8.33	&	0.08	\\
54309.90	&	8.21	&	0.07	\\
54309.97	&	8.68	&	0.09	\\
54310.04	&	8.58	&	0.09	\\
54310.11	&	8.65	&	0.09	\\
54310.18	&	8.61	&	0.09	\\
54310.25	&	8.31	&	0.08	\\
54310.32	&	9.13	&	0.11	\\
54310.39	&	8.73	&	0.09	\\
54310.47	&	9.05	&	0.11	\\
54310.61	&	9.08	&	0.11	\\
54310.68	&	8.04	&	0.07	\\
54310.75	&	9.22	&	0.12	\\
54310.82	&	11.30	&	0.30	\\
54310.89	&	9.57	&	0.14	\\
54310.96	&	7.92	&	0.06	\\
54311.17	&	8.71	&	0.09	\\
54311.31	&	8.49	&	0.08	\\
54311.38	&	9.26	&	0.12	\\
54311.45	&	7.49	&	0.05	\\
54311.52	&	8.81	&	0.10	\\
54311.59	&	11.19	&	0.28	\\
54311.74	&	14.67	&	1.38	\\
54311.81	&	8.37	&	0.08	\\
54311.88	&	5.97	&	0.03	\\
54312.02	&	8.84	&	0.10	\\
54312.09	&	7.17	&	0.05	\\
54312.16	&	8.24	&	0.07	\\
54312.23	&	6.06	&	0.03	\\
54312.30	&	6.15	&	0.03	\\
54312.37	&	7.36	&	0.05	\\
54312.44	&	6.40	&	0.03	\\
54312.51	&	5.95	&	0.03	\\
54312.58	&	5.97	&	0.03	\\
54312.65	&	5.91	&	0.03	\\
54312.72	&	5.90	&	0.03	\\
54312.79	&	6.40	&	0.03	\\
54312.86	&	6.40	&	0.03	\\
54312.93	&	6.02	&	0.03	\\
54313.29	&	9.20	&	0.11	\\
54313.36	&	6.45	&	0.03	\\
54313.43	&	6.29	&	0.03	\\
54313.50	&	7.06	&	0.04	\\
54313.57	&	7.43	&	0.05	\\
54313.78	&	7.84	&	0.06	\\
54313.85	&	8.40	&	0.08	\\
54313.92	&	7.28	&	0.05	\\
54313.99	&	8.49	&	0.08	\\
54314.06	&	7.90	&	0.06	\\
54314.13	&	8.64	&	0.09	\\
54314.20	&	7.08	&	0.04	\\
54314.27	&	11.04	&	0.26	\\
54314.35	&	8.41	&	0.08	\\
54314.42	&	8.35	&	0.08	\\
54314.49	&	7.53	&	0.05	\\
54314.63	&	7.54	&	0.05	\\
54314.70	&	7.92	&	0.06	\\
54314.77	&	8.38	&	0.08	\\
54314.84	&	7.59	&	0.06	\\
54314.91	&	8.91	&	0.10	\\
54314.98	&	8.69	&	0.09	\\
54315.05	&	7.53	&	0.05	\\
    \end{tabular}
\end{minipage}
\begin{minipage}[t]{0.333\linewidth}
	\begin{tabular}[t]{lcc}
Date-Obs MJD & SMEI Mag & SMEI Error \\
\hline
54315.12	&	7.93	&	0.06	\\
54315.19	&	9.62	&	0.14	\\
54315.40	&	15.13	&	1.71	\\
54315.54	&	8.31	&	0.08	\\
54315.61	&	8.73	&	0.09	\\
54315.76	&	10.90	&	0.25	\\
54315.83	&	9.13	&	0.11	\\
54315.90	&	8.61	&	0.09	\\
54315.97	&	9.68	&	0.14	\\
54316.04	&	8.39	&	0.08	\\
54316.11	&	9.14	&	0.11	\\
54316.18	&	8.06	&	0.07	\\
54316.25	&	7.62	&	0.06	\\
54316.32	&	8.47	&	0.08	\\
54316.39	&	8.95	&	0.10	\\
54316.53	&	8.38	&	0.08	\\
54316.60	&	8.39	&	0.08	\\
54316.67	&	8.67	&	0.09	\\
54316.74	&	8.33	&	0.08	\\
54316.81	&	8.64	&	0.09	\\
54316.88	&	8.35	&	0.08	\\
54317.10	&	9.54	&	0.13	\\
54317.31	&	10.29	&	0.19	\\
54317.38	&	10.89	&	0.25	\\
54317.45	&	8.62	&	0.09	\\
54317.52	&	7.82	&	0.06	\\
54317.59	&	9.60	&	0.14	\\
54317.66	&	8.23	&	0.07	\\
54317.73	&	8.28	&	0.08	\\
54317.80	&	7.75	&	0.06	\\
54317.87	&	9.46	&	0.13	\\
54317.94	&	8.10	&	0.07	\\
54318.01	&	7.95	&	0.06	\\
54318.08	&	8.17	&	0.07	\\
54318.15	&	7.26	&	0.05	\\
54318.22	&	11.12	&	0.27	\\
54318.30	&	8.48	&	0.08	\\
54318.44	&	8.41	&	0.08	\\
54318.51	&	13.05	&	0.66	\\
54318.58	&	9.05	&	0.11	\\
54318.65	&	8.41	&	0.08	\\
54318.72	&	8.69	&	0.09	\\
54318.79	&	11.58	&	0.34	\\
54318.86	&	8.96	&	0.10	\\
54318.93	&	7.52	&	0.05	\\
54319.00	&	8.52	&	0.08	\\
54319.07	&	9.98	&	0.16	\\
54319.21	&	8.34	&	0.08	\\
54319.35	&	8.52	&	0.08	\\
54319.42	&	7.46	&	0.05	\\
54319.49	&	8.15	&	0.07	\\
54319.57	&	11.33	&	0.30	\\
54319.64	&	8.42	&	0.08	\\
54319.71	&	7.40	&	0.05	\\
54319.78	&	6.81	&	0.04	\\
54319.85	&	8.53	&	0.08	\\
54319.92	&	8.36	&	0.08	\\
54319.99	&	8.22	&	0.07	\\
54320.27	&	8.90	&	0.10	\\
54320.34	&	8.23	&	0.07	\\
54320.41	&	8.18	&	0.07	\\
54320.48	&	7.79	&	0.06	\\
54320.55	&	8.14	&	0.07	\\
54320.62	&	8.51	&	0.08	\\
54320.69	&	8.67	&	0.09	\\
54320.76	&	8.32	&	0.08	\\
54320.84	&	9.45	&	0.13	\\
54320.91	&	8.72	&	0.09	\\
54321.12	&	8.64	&	0.09	\\
54321.19	&	7.80	&	0.06	\\
    \end{tabular}
\end{minipage}
\hrule
\end{table}


\begin{table}
\scriptsize
\caption{SMEI light curve data for Nova V598 Pup continued}
\hrule
\begin{minipage}[t]{0.333\linewidth}
	\begin{tabular}[t]{lcc}
Date-Obs MJD & SMEI Mag & SMEI Error \\
\hline
54321.26	&	8.26	&	0.07	\\
54321.47	&	8.56	&	0.09	\\
54321.54	&	8.73	&	0.09	\\
54321.61	&	9.40	&	0.13	\\
54321.68	&	9.08	&	0.11	\\
54321.75	&	9.88	&	0.16	\\
54321.82	&	9.48	&	0.13	\\
54321.89	&	6.84	&	0.04	\\
54321.96	&	9.38	&	0.12	\\
54322.10	&	8.04	&	0.07	\\
54322.18	&	11.40	&	0.31	\\
54322.25	&	8.53	&	0.08	\\
54322.32	&	9.09	&	0.11	\\
54322.39	&	7.61	&	0.06	\\
54322.46	&	6.88	&	0.04	\\
54322.53	&	7.10	&	0.04	\\
54322.60	&	8.16	&	0.07	\\
54322.67	&	7.18	&	0.05	\\
54322.74	&	7.73	&	0.06	\\
54322.95	&	7.65	&	0.06	\\
54323.23	&	8.04	&	0.07	\\
54323.37	&	9.96	&	0.16	\\
54323.45	&	9.68	&	0.14	\\
54323.59	&	8.15	&	0.07	\\
54323.66	&	6.88	&	0.04	\\
54323.73	&	9.08	&	0.11	\\
54323.80	&	7.46	&	0.05	\\
54323.87	&	7.86	&	0.06	\\
54323.94	&	7.62	&	0.06	\\
54324.01	&	7.97	&	0.07	\\
54324.08	&	8.34	&	0.08	\\
54324.15	&	7.64	&	0.06	\\
54324.29	&	7.98	&	0.07	\\
54324.57	&	7.95	&	0.06	\\
54324.64	&	7.37	&	0.05	\\
54324.72	&	7.43	&	0.05	\\
54324.79	&	7.56	&	0.05	\\
    \end{tabular}
\end{minipage}
\begin{minipage}[t]{0.333\linewidth}
	\begin{tabular}[t]{lcc}
Date-Obs MJD & SMEI Mag & SMEI Error \\
\hline
54324.86	&	8.51	&	0.08	\\
54324.93	&	9.02	&	0.11	\\
54325.14	&	7.23	&	0.05	\\
54325.21	&	7.67	&	0.06	\\
54325.28	&	10.04	&	0.17	\\
54325.35	&	9.85	&	0.15	\\
54325.42	&	7.43	&	0.05	\\
54325.63	&	8.67	&	0.09	\\
54325.70	&	8.47	&	0.08	\\
54325.77	&	7.99	&	0.07	\\
54325.84	&	9.75	&	0.15	\\
54325.91	&	8.30	&	0.08	\\
54325.98	&	8.70	&	0.09	\\
54326.06	&	9.49	&	0.13	\\
54326.20	&	6.47	&	0.03	\\
54326.27	&	8.20	&	0.07	\\
54326.34	&	10.41	&	0.20	\\
54326.41	&	8.53	&	0.08	\\
54326.48	&	9.59	&	0.14	\\
54326.55	&	7.47	&	0.05	\\
54326.62	&	9.28	&	0.12	\\
54326.76	&	8.02	&	0.07	\\
54326.83	&	8.39	&	0.08	\\
54326.90	&	9.02	&	0.11	\\
54326.97	&	8.01	&	0.07	\\
54327.25	&	8.02	&	0.07	\\
54327.33	&	9.54	&	0.13	\\
54327.40	&	9.27	&	0.12	\\
54327.47	&	7.47	&	0.05	\\
54327.54	&	9.75	&	0.15	\\
54327.61	&	9.44	&	0.13	\\
54327.68	&	9.20	&	0.11	\\
54327.75	&	7.76	&	0.06	\\
54327.82	&	9.19	&	0.11	\\
54327.89	&	9.00	&	0.10	\\
54327.96	&	9.17	&	0.11	\\
54328.03	&	8.78	&	0.09	\\
    \end{tabular}
\end{minipage}
\begin{minipage}[t]{0.333\linewidth}
	\begin{tabular}[t]{lcc}
Date-Obs MJD & SMEI Mag & SMEI Error \\
\hline
54328.10	&	8.19	&	0.07	\\
54328.17	&	8.40	&	0.08	\\
54328.24	&	8.58	&	0.09	\\
54328.31	&	8.39	&	0.08	\\
54328.45	&	10.60	&	0.22	\\
54328.67	&	9.57	&	0.14	\\
54328.74	&	9.56	&	0.13	\\
54328.81	&	8.16	&	0.07	\\
54328.88	&	8.27	&	0.08	\\
54328.95	&	9.24	&	0.12	\\
54329.02	&	7.62	&	0.06	\\
54329.09	&	8.58	&	0.09	\\
54329.16	&	8.87	&	0.10	\\
54329.23	&	9.06	&	0.11	\\
54329.37	&	8.51	&	0.08	\\
54329.44	&	10.23	&	0.18	\\
54329.51	&	8.49	&	0.08	\\
54329.58	&	9.00	&	0.10	\\
54329.65	&	8.43	&	0.08	\\
54329.72	&	9.09	&	0.11	\\
54329.79	&	9.20	&	0.11	\\
54329.87	&	11.10	&	0.27	\\
54329.94	&	8.55	&	0.09	\\
54330.01	&	8.21	&	0.07	\\
54330.08	&	11.81	&	0.38	\\
54330.15	&	8.35	&	0.08	\\
54330.29	&	8.54	&	0.09	\\
54330.36	&	8.33	&	0.08	\\
54330.43	&	7.37	&	0.05	\\
54330.50	&	7.79	&	0.06	\\
54330.57	&	9.18	&	0.11	\\
54330.64	&	8.78	&	0.09	\\
54330.71	&	7.83	&	0.06	\\
54330.85	&	9.85	&	0.15	\\
54330.99	&	8.70	&	0.09	\\
    \end{tabular}
\end{minipage}
\hrule
\end{table}


\begin{table}
\scriptsize
\caption{SMEI light curve data for Nova KT Eri}
\hrule
\begin{minipage}[t]{0.333\linewidth}
	\begin{tabular}[t]{lcc}
Date-Obs MJD & SMEI Mag & SMEI Error \\
\hline
55146.99	&	9.51	&	0.15	\\
55148.12	&	8.44	&	0.09	\\
55148.19	&	8.05	&	0.08	\\
55148.26	&	7.83	&	0.07	\\
55148.33	&	7.50	&	0.06	\\
55148.40	&	7.28	&	0.05	\\
55148.47	&	7.01	&	0.05	\\
55148.54	&	6.84	&	0.04	\\
55148.62	&	6.67	&	0.04	\\
55148.69	&	6.52	&	0.04	\\
55148.76	&	6.35	&	0.04	\\
55148.83	&	6.17	&	0.03	\\
55148.90	&	5.94	&	0.03	\\
55148.97	&	6.02	&	0.03	\\
55149.25	&	5.78	&	0.03	\\
55149.32	&	5.74	&	0.03	\\
55149.39	&	5.61	&	0.03	\\
55149.46	&	5.53	&	0.02	\\
55149.53	&	5.47	&	0.02	\\
55149.60	&	5.45	&	0.02	\\
55149.67	&	5.42	&	0.02	\\
55149.74	&	5.44	&	0.02	\\
55149.81	&	5.47	&	0.02	\\
55149.89	&	5.49	&	0.02	\\
55149.96	&	5.59	&	0.02	\\
55150.03	&	5.62	&	0.03	\\
55150.10	&	5.65	&	0.03	\\
55150.17	&	5.80	&	0.03	\\
55150.24	&	5.85	&	0.03	\\
55150.45	&	5.90	&	0.03	\\
55150.52	&	5.88	&	0.03	\\
55150.59	&	5.94	&	0.03	\\
55150.66	&	5.95	&	0.03	\\
55150.73	&	5.98	&	0.03	\\
55150.80	&	6.01	&	0.03	\\
55151.08	&	6.19	&	0.03	\\
55151.16	&	6.19	&	0.03	\\
55151.23	&	6.24	&	0.03	\\
55151.30	&	6.23	&	0.03	\\
55151.37	&	6.31	&	0.03	\\
55151.44	&	6.34	&	0.03	\\
55151.51	&	6.33	&	0.03	\\
55151.58	&	6.34	&	0.03	\\
55151.65	&	6.35	&	0.04	\\
55151.72	&	6.36	&	0.04	\\
55151.79	&	6.37	&	0.04	\\
55151.86	&	6.38	&	0.04	\\
55151.93	&	6.32	&	0.03	\\
55152.00	&	6.43	&	0.04	\\
55152.07	&	6.44	&	0.04	\\
55152.14	&	6.47	&	0.04	\\
55152.21	&	6.44	&	0.04	\\
55152.28	&	6.52	&	0.04	\\
    \end{tabular}
\end{minipage}
\begin{minipage}[t]{0.333\linewidth}
	\begin{tabular}[t]{lcc}
Date-Obs MJD & SMEI Mag & SMEI Error \\
\hline
55152.35	&	6.50	&	0.04	\\
55152.43	&	6.53	&	0.04	\\
55152.50	&	6.56	&	0.04	\\
55152.57	&	6.60	&	0.04	\\
55152.64	&	6.60	&	0.04	\\
55152.71	&	6.61	&	0.04	\\
55152.78	&	6.58	&	0.04	\\
55152.85	&	6.52	&	0.04	\\
55152.92	&	6.65	&	0.04	\\
55153.27	&	6.69	&	0.04	\\
55153.34	&	6.65	&	0.04	\\
55153.41	&	6.72	&	0.04	\\
55153.48	&	6.75	&	0.04	\\
55153.55	&	6.72	&	0.04	\\
55153.63	&	6.82	&	0.04	\\
55153.70	&	6.88	&	0.04	\\
55153.77	&	6.93	&	0.05	\\
55153.84	&	6.93	&	0.05	\\
55153.91	&	6.98	&	0.05	\\
55153.98	&	7.01	&	0.05	\\
55154.05	&	7.04	&	0.05	\\
55154.12	&	7.03	&	0.05	\\
55154.19	&	7.07	&	0.05	\\
55154.26	&	7.11	&	0.05	\\
55154.33	&	7.14	&	0.05	\\
55154.40	&	7.20	&	0.05	\\
55154.75	&	7.20	&	0.05	\\
55154.82	&	7.21	&	0.05	\\
55154.90	&	7.27	&	0.05	\\
55154.97	&	7.29	&	0.05	\\
55155.04	&	7.32	&	0.05	\\
55155.11	&	7.30	&	0.05	\\
55155.18	&	7.35	&	0.06	\\
55155.25	&	7.28	&	0.05	\\
55155.32	&	7.26	&	0.05	\\
55155.39	&	7.22	&	0.05	\\
55155.53	&	7.22	&	0.05	\\
55155.60	&	7.23	&	0.05	\\
55155.67	&	7.25	&	0.05	\\
55155.74	&	7.30	&	0.05	\\
55155.81	&	7.36	&	0.06	\\
55155.88	&	7.36	&	0.06	\\
55155.95	&	7.35	&	0.06	\\
55156.02	&	7.35	&	0.06	\\
55156.09	&	7.37	&	0.06	\\
55156.17	&	7.34	&	0.05	\\
55156.24	&	7.47	&	0.06	\\
55156.31	&	7.52	&	0.06	\\
55156.38	&	7.50	&	0.06	\\
55156.52	&	7.55	&	0.06	\\
55156.59	&	7.51	&	0.06	\\
55156.66	&	7.59	&	0.06	\\
55156.73	&	7.60	&	0.06	\\
    \end{tabular}
\end{minipage}
\begin{minipage}[t]{0.333\linewidth}
	\begin{tabular}[t]{lcc}
Date-Obs MJD & SMEI Mag & SMEI Error \\
\hline
55156.80	&	7.66	&	0.06	\\
55156.87	&	7.62	&	0.06	\\
55156.94	&	7.63	&	0.06	\\
55157.01	&	7.59	&	0.06	\\
55157.08	&	7.59	&	0.06	\\
55157.15	&	7.58	&	0.06	\\
55157.22	&	7.56	&	0.06	\\
55157.37	&	7.62	&	0.06	\\
55157.44	&	7.56	&	0.06	\\
55157.51	&	7.62	&	0.06	\\
55157.58	&	7.65	&	0.06	\\
55157.65	&	7.66	&	0.06	\\
55157.72	&	7.56	&	0.06	\\
55157.79	&	7.62	&	0.06	\\
55157.86	&	7.68	&	0.06	\\
55158.14	&	7.77	&	0.07	\\
55158.21	&	7.84	&	0.07	\\
55158.28	&	7.78	&	0.07	\\
55158.35	&	7.76	&	0.07	\\
55158.42	&	7.83	&	0.07	\\
55158.56	&	7.84	&	0.07	\\
55158.64	&	7.78	&	0.07	\\
55158.71	&	7.96	&	0.07	\\
55158.78	&	7.83	&	0.07	\\
55158.85	&	7.92	&	0.07	\\
55158.92	&	7.95	&	0.07	\\
55158.99	&	8.01	&	0.07	\\
55159.06	&	8.03	&	0.08	\\
55159.13	&	8.03	&	0.08	\\
55159.20	&	8.04	&	0.08	\\
55159.27	&	8.05	&	0.08	\\
55159.34	&	8.14	&	0.08	\\
55159.41	&	8.19	&	0.08	\\
55159.62	&	8.26	&	0.08	\\
55159.69	&	8.25	&	0.08	\\
55159.76	&	8.22	&	0.08	\\
55159.84	&	8.26	&	0.08	\\
55159.91	&	8.23	&	0.08	\\
55159.98	&	8.17	&	0.08	\\
55160.26	&	8.21	&	0.08	\\
55160.33	&	8.24	&	0.08	\\
55160.40	&	8.23	&	0.08	\\
55160.47	&	8.20	&	0.08	\\
55160.54	&	8.23	&	0.08	\\
55160.61	&	8.35	&	0.09	\\
55160.68	&	8.31	&	0.09	\\
55160.82	&	8.35	&	0.09	\\
55160.89	&	8.29	&	0.08	\\
55160.96	&	8.29	&	0.08	\\
55161.03	&	8.23	&	0.08	\\
55161.11	&	8.35	&	0.09	\\
55161.18	&	8.33	&	0.09	\\
55162.23	&	8.27	&	0.08	\\
    \end{tabular}
\end{minipage}
\hrule
\end{table}

\end{document}